\documentclass[preprint,cm]{sigplanconf}
\usepackage{caption}
\DeclareCaptionType{copyrightbox}

\makeatletter
\@ifundefined{lhs2tex.lhs2tex.sty.read}%
  {\@namedef{lhs2tex.lhs2tex.sty.read}{}%
   \newcommand\SkipToFmtEnd{}%
   \newcommand\EndFmtInput{}%
   \long\def\SkipToFmtEnd#1\EndFmtInput{}%
  }\SkipToFmtEnd

\newcommand\ReadOnlyOnce[1]{\@ifundefined{#1}{\@namedef{#1}{}}\SkipToFmtEnd}
\usepackage{amstext}
\usepackage{amssymb}
\usepackage{stmaryrd}
\DeclareFontFamily{OT1}{cmtex}{}
\DeclareFontShape{OT1}{cmtex}{m}{n}
  {<5><6><7><8>cmtex8
   <9>cmtex9
   <10><10.95><12><14.4><17.28><20.74><24.88>cmtex10}{}
\DeclareFontShape{OT1}{cmtex}{m}{it}
  {<-> ssub * cmtt/m/it}{}

\DeclareFontShape{OT1}{cmtt}{bx}{n}
  {<5><6><7><8>cmtt8
   <9>cmbtt9
   <10><10.95><12><14.4><17.28><20.74><24.88>cmbtt10}{}
\DeclareFontShape{OT1}{cmtex}{bx}{n}
  {<-> ssub * cmtt/bx/n}{}
\newcommand{\Conid}[1]{\mathit{#1}}
\newcommand{\Varid}[1]{\mathit{#1}}
\newcommand{\anonymous}{\kern0.06em \vbox{\hrule\@width.5em}}

\usepackage{polytable}

\@ifundefined{mathindent}%
  {\newdimen\mathindent\mathindent\leftmargini}%
  {}%

\def\resethooks{%
  \global\let\SaveRestoreHook\empty
  \global\let\ColumnHook\empty}
\newcommand*{\savecolumns}[1][default]%
  {\g@addto@macro\SaveRestoreHook{\savecolumns[#1]}}
\newcommand*{\restorecolumns}[1][default]%
  {\g@addto@macro\SaveRestoreHook{\restorecolumns[#1]}}
\newcommand*{\aligncolumn}[2]%
  {\g@addto@macro\ColumnHook{\column{#1}{#2}}}

\resethooks

\newcommand{\onelinecommentchars}{\quad-{}- }
\newcommand{\commentbeginchars}{\enskip\{-}
\newcommand{\commentendchars}{-\}\enskip}

\newcommand{\visiblecomments}{%
  \let\onelinecomment=\onelinecommentchars
  \let\commentbegin=\commentbeginchars
  \let\commentend=\commentendchars}

\newcommand{\invisiblecomments}{%
  \let\onelinecomment=\empty
  \let\commentbegin=\empty
  \let\commentend=\empty}

\visiblecomments

\newlength{\blanklineskip}
\newcommand{\hsindent}[1]{\quad}% default is fixed indentation
\let\hspre\empty
\let\hspost\empty

\EndFmtInput
\makeatother
\ReadOnlyOnce{polycode.fmt}%
\makeatletter

\newcommand{\hsnewpar}[1]%
  {{\parskip=0pt\parindent=0pt\par\vskip #1\noindent}}

\newcommand{\hscodestyle}{}

\newcommand{\sethscode}[1]%
  {\expandafter\let\expandafter\hscode\csname #1\endcsname
   \expandafter\let\expandafter\endhscode\csname end#1\endcsname}

  {\par\noindent
   \advance\leftskip\mathindent
   \hscodestyle
   \let\\=\@normalcr
   \let\hspre\(\let\hspost\)%
   \pboxed}%
  {\endpboxed\)%
   \par\noindent
   \ignorespacesafterend}

  {\hsnewpar\abovedisplayskip
   \advance\leftskip\mathindent
   \hscodestyle
   \let\hspre\(\let\hspost\)%
   \pboxed}%
  {\endpboxed%
   \hsnewpar\belowdisplayskip
   \ignorespacesafterend}

  {\hsnewpar\abovedisplayskip
   \advance\leftskip\mathindent
   \hscodestyle
   \let\\=\@normalcr
   \(\pboxed}%
  {\endpboxed\)%
   \hsnewpar\belowdisplayskip
   \ignorespacesafterend}

\newcommand{\plainhs}{\sethscode{plainhscode}}

\plainhs

  {\hsnewpar\abovedisplayskip
   \advance\leftskip\mathindent
   \hscodestyle
   \let\\=\@normalcr
   \(\parray}%
  {\endparray\)%
   \hsnewpar\belowdisplayskip
   \ignorespacesafterend}

  {\parray}{\endparray}

  {\(\parray}{\endparray\)}

\def\codeframewidth{\arrayrulewidth}
\RequirePackage{calc}

  {\parskip=\abovedisplayskip\par\noindent
   \hscodestyle
   \arrayrulewidth=\codeframewidth
   \tabular{@{}|p{\linewidth-2\arraycolsep-2\arrayrulewidth-2pt}|@{}}%
   \hline\framedhslinecorrect\\{-1.5ex}%
   \let\endoflinesave=\\
   \let\\=\@normalcr
   \(\pboxed}%
  {\endpboxed\)%
   \framedhslinecorrect\endoflinesave{.5ex}\hline
   \endtabular
   \parskip=\belowdisplayskip\par\noindent
   \ignorespacesafterend}

\newcommand{\framedhslinecorrect}[2]%
  {#1[#2]}

  {\(\def\column##1##2{}%
   \let\>\undefined\let\<\undefined\let\\\undefined
   \newcommand\>[1][]{}\newcommand\<[1][]{}\newcommand\\[1][]{}%
   \def\fromto##1##2##3{##3}%
   }{\) }%

  {\let\orighscode=\hscode
   \let\origendhscode=\endhscode
   \def\endhscode{\def\hscode{\endgroup\def\@currenvir{hscode}\\}\begingroup}
   \orighscode\def\hscode{\endgroup\def\@currenvir{hscode}}}%
  {\origendhscode
   \global\let\hscode=\orighscode
   \global\let\endhscode=\origendhscode}%

\makeatother
\EndFmtInput
\ReadOnlyOnce{agda.fmt}%

\RequirePackage[T1]{fontenc}
\RequirePackage[utf8x]{inputenc}
\RequirePackage{ucs}
\RequirePackage{amsfonts}

\DeclareUnicodeCharacter{737}{\textsuperscript{l}}
\DeclareUnicodeCharacter{8718}{\ensuremath{\blacksquare}}
\DeclareUnicodeCharacter{8759}{::}
\DeclareUnicodeCharacter{9669}{\ensuremath{\triangleleft}}
\DeclareUnicodeCharacter{8799}{\ensuremath{\stackrel{\scriptscriptstyle ?}{=}}}
\DeclareUnicodeCharacter{10214}{\ensuremath{\llbracket}}
\DeclareUnicodeCharacter{10215}{\ensuremath{\rrbracket}}

\renewcommand\Varid[1]{\mathord{\textsf{#1}}}
\let\Conid\Varid
\newcommand\Keyword[1]{\textsf{\textbf{#1}}}
\EndFmtInput

\usepackage{amsmath,amsthm}
\usepackage{semantic}
\usepackage{subfig}

\usepackage{fancyvrb}
\DefineVerbatimEnvironment{code}{Verbatim}{}
  
\newcommand{\map}{\textrm{map}\,}
\newcommand{\id}{\textrm{id}}

\newcommand{\some}{\textrm{some}\;}

\newcommand{\nat}{\textrm{nat}\;}
\newcommand{\nilE}{\textrm{nilE}\;}

\mathchardef\mhyphen="2D
\mathchardef\textpipe\mathcode`|

\DeclareUnicodeCharacter{948}{$\delta$}
\DeclareUnicodeCharacter{955}{$\lambda$}
\DeclareUnicodeCharacter{964}{$\tau$}
\DeclareUnicodeCharacter{8611}{$\rightarrowtail$}
\DeclareUnicodeCharacter{8788}{$:=$}
\DeclareUnicodeCharacter{10230}{$\longrightarrow$}

\title{Modular Type-Safety Proofs using Dependant Types}
\authorinfo{Christopher Schwaab\and Jeremy G. Siek}
           {University of Colorado at Boulder}
           {\{schwaab,jeremy.siek\}@colorado.edu}
\date{}

\begin{document}
\maketitle

\begin{abstract}

While methods of code abstraction and reuse are widespread and well
researched, methods of proof abstraction and reuse are still emerging.
We consider the use of dependent types for this purpose, introducing a
completely mechanical approach to proof composition.  We show that
common techniques for abstracting algorithms over data structures
naturally translate to abstractions over proofs.  We first introduce a
language composed of a series of smaller language components tied
together by standard techniques from Malcom~\cite{malcom}.  We proceed
by giving proofs of type preservation for each language component and
show that the basic ideas used in composing the syntactic data
structures can be applied to their semantics as well.

\end{abstract}

\section{Introduction}

The POPLmark challenge is a set of common programming language
problems meant to test the utility of modern proof assistants and
techniques for mechanized metatheory. In response to this challenge,
significant strides have been made in making it easier to mechanize
the metatheory of programming languages, especially regarding variable
binding~\cite{metatheory for the masses}. However, little progress has
been made in the direction of modularity: it is still difficult to
separately develop the definitions and meta-theory of language
fragments and then link the fragments together to obtain the
definitions and meta-theory for a language composed of such fragments.

% The following is rather tangential and probably not of interest
% to the audience. -Jeremy
% Unlike in mathematics where proofs are rich and naturally extend
% to new objects when new abstractions are discovered, many common language
% theoretic properties such as type-safety and termination are direct,
% frequently requring several similar cases. 
% While it is fortunate that language
% features are often developed in a distributed fashion whereby extensions can
% be developed atop recycled ideas---proofs cannot and well understood
% properties must be reshown following each new discovery. Current efforts
% have made great strides in reusability, submissions to the POPLmark challenge
% leverage a broad range of techniques to increase
% automation\cite{metatheory for the masses}; despite this, existing proof
% assistants are unable to automatically derive such well known properties
% burdening the implementor with another dry set of case analyses.

Dependent types have formed the foundation of a broad and rich range of
type systems that allow values and types to be freely mixed.  Programmers can
express propositions as types viewed as sets, and proofs as objects viewed
as inhabitants of those sets.  This style of theorem proving suggests the use of
familiar engineering abstractions as general solutions to questions
about theorem proving.  Rather than relying on semi-automated proof search
such as Coq's Ltac we propose a method of proof composition using simple
abstractions whereby components are defined piecewise and ``tied'' together
at the end using a wrapper datatype acting as a tagged union.

The method of language definition used is iterative.
Components are defined separately from one another and are composable along
with their proofs.  Thus we would like for separate language designers to be
able to reuse one anothers' work without the need for sophisticated proof
search algorithms or with effort spent copying and pasting terms.

The language we present is one of simple expressons using Agda as the
implementation language and proof assistant. We begin by defining a
series of language syntaxes for sums, options, and arrays.  We chose
to include arrays because they not only can result in runtime errors
requiring the inclusion of the $Option$ type but like addition, they
use the natural numbers, forcing consideration of how value types can
be shared across otherwise isolated components.  We continue by
defining evaluation semantics and typing rules. The language is
defined piecewise, each component is built in isolation alongside a
proof of type preservation. We conclude with a presentation of how
these components can be composed and a proof of type preservation for
the combined language can be immediately derived from the
componentwise proofs.  The motivation for our technique is drawn from
a solution to the expression problem where languages are defined as
the disjoint sum of smaller languages by removing explicit recursion.
We show that this idea can be recast from types and terms, to proofs.

\section{A Review of the Expression Problem}
When modeling a problem with a functional flavor often the natural solution
emerges as several recursive cases handled by some helper functions.
The expression problem states that this type of solution presents us with a
choice: we may ordain our data structure forever unchanging, making it
easy to add new functions without changing the program; or we may leave our
data structure open, making it difficult to extend the original program
with new functions.

While many solutions to the expression problem have been proposed over the
years, here we make use of the method described by Malcom~\cite{malcom} which
generalizes recursion operators such as fold, from lists to polynomial types.
The problem we encounter arises as a result of algebraic data types being
\emph{closed}: once the type has been declared, no new constructors for the
type may be added without amending the original declaration and the solution
presented lies at the heart of our work.  The idea is simply to remove
immediate recursion and split a monolithic datatype into components to be later
collected under the umbrella of a tagged union.

Throughout this paper we will work with a simple evaluator over natural numbers
and basic arithmetic operators; in Agda we might first consider
\begin{hscode}\SaveRestoreHook
\column{B}{@{}>{\hspre}l<{\hspost}@{}}%
\column{3}{@{}>{\hspre}l<{\hspost}@{}}%
\column{E}{@{}>{\hspre}l<{\hspost}@{}}%
\>[B]{}\Keyword{data}\;\Conid{Expr+}\;\mathbin{:}\;\Conid{Set}\;\Keyword{where}{}\<[E]%
\\
\>[B]{}\hsindent{3}{}\<[3]%
\>[3]{}\Varid{atom}\;\mathbin{:}\;\Conid{ℕ}\;\Varid{→}\;\Conid{Expr+}{}\<[E]%
\\
\>[B]{}\hsindent{3}{}\<[3]%
\>[3]{}\Varid{\char95 +\char95 }\;\mathbin{:}\;\Conid{Expr+}\;\Varid{→}\;\Conid{Expr+}\;\Varid{→}\;\Conid{Expr+}{}\<[E]%
\ColumnHook
\end{hscode}\resethooks
This definition has the advantage of being direct and simple, however
a problem lies within the explicit recursion; notice that when later
extending expressions with arrays and option types we can make no reuse
of {\tt Expr+} due to the closed nature of algebraic data types.
To extend {\tt Expr+} we must define a whole new data type,
as in the following definition of {\tt MonolithicExpr}.

\begin{hscode}\SaveRestoreHook
\column{B}{@{}>{\hspre}l<{\hspost}@{}}%
\column{3}{@{}>{\hspre}l<{\hspost}@{}}%
\column{5}{@{}>{\hspre}l<{\hspost}@{}}%
\column{20}{@{}>{\hspre}l<{\hspost}@{}}%
\column{E}{@{}>{\hspre}l<{\hspost}@{}}%
\>[B]{}\Keyword{data}\;\Conid{MonolithicExpr}\;\mathbin{:}\;\Conid{Set}\;\Keyword{where}{}\<[E]%
\\
\>[B]{}\hsindent{3}{}\<[3]%
\>[3]{}\Varid{atom}\;\mathbin{:}\;\Conid{ℕ}\;\Varid{→}\;\Conid{MonolithicExpr}{}\<[E]%
\\
\>[B]{}\hsindent{3}{}\<[3]%
\>[3]{}\Varid{esome}\;\mathbin{:}\;\Conid{MonolithicExpr}\;\Varid{→}\;\Conid{MonolithicExpr}{}\<[E]%
\\
\>[B]{}\hsindent{3}{}\<[3]%
\>[3]{}\Varid{enone}\;\mathbin{:}\;\Conid{MonolithicExpr}{}\<[E]%
\\
\>[B]{}\hsindent{3}{}\<[3]%
\>[3]{}\Varid{nil}\;[\mskip1.5mu \mskip1.5mu]\;\mathbin{:}\;\Conid{MonolithicExpr}{}\<[E]%
\\
\>[B]{}\hsindent{3}{}\<[3]%
\>[3]{}\Varid{\char95 !!\char95 }\;\mathbin{:}\;\Conid{MonolithicExpr}\;\Varid{→}\;\Conid{MonolithicExpr}\;{}\<[E]%
\\
\>[3]{}\hsindent{2}{}\<[5]%
\>[5]{}\Varid{→}\;\Conid{MonolithicExpr}{}\<[E]%
\\
\>[B]{}\hsindent{3}{}\<[3]%
\>[3]{}\anonymous \;[\mskip1.5mu \anonymous \mskip1.5mu]\;\Varid{≔\char95 }\;\mathbin{:}\;\Conid{MonolithicExpr}\;\Varid{→}\;\Conid{MonolithicExpr}\;{}\<[E]%
\\
\>[3]{}\hsindent{2}{}\<[5]%
\>[5]{}\Varid{→}\;\Conid{MonolithicExpr}\;\Varid{→}\;\Conid{MonolithicExpr}{}\<[E]%
\\
\>[B]{}\hsindent{3}{}\<[3]%
\>[3]{}\Varid{\char95 +\char95 }\;\mathbin{:}\;\Conid{MonolithicExpr}\;\Varid{→}\;\Conid{MonolithicExpr}\;{}\<[E]%
\\
\>[3]{}\hsindent{2}{}\<[5]%
\>[5]{}\Varid{→}\;\Conid{MonolithicExpr}{}\<[E]%
\\
\>[B]{}\Varid{fromExpr+}\;\mathbin{:}\;\Conid{Expr+}\;\Varid{→}\;\Conid{MonolithicExpr}{}\<[E]%
\\
\>[B]{}\Varid{fromExpr+}\;(\Varid{atom}\;\Varid{n})\;\mathrel{=}\;\Varid{atom}\;\Varid{n}{}\<[E]%
\\
\>[B]{}\Varid{fromExpr+}\;(\Varid{n}\;\Varid{+}\;\Varid{m})\;{}\<[20]%
\>[20]{}\mathrel{=}\;\Varid{fromExpr+}\;\Varid{n}\;\Varid{+}\;\Varid{fromExpr+}\;\Varid{m}{}\<[E]%
\ColumnHook
\end{hscode}\resethooks
Suppose instead we begin with polymorphic definitions such as the
following.
\begin{hscode}\SaveRestoreHook
\column{B}{@{}>{\hspre}l<{\hspost}@{}}%
\column{3}{@{}>{\hspre}l<{\hspost}@{}}%
\column{E}{@{}>{\hspre}l<{\hspost}@{}}%
\>[B]{}\Keyword{data}\;\Conid{Expr+₂}\;(\Conid{A}\;\mathbin{:}\;\Conid{Set})\;\mathbin{:}\;\Conid{Set}\;\Keyword{where}{}\<[E]%
\\
\>[B]{}\hsindent{3}{}\<[3]%
\>[3]{}\Varid{\char95 +\char95 }\;\mathbin{:}\;\Conid{A}\;\Varid{→}\;\Conid{A}\;\Varid{→}\;\Conid{Expr+₂}\;\Conid{A}{}\<[E]%
\\
\>[B]{}\Keyword{data}\;\Conid{Expr}\;[\mskip1.5mu \mskip1.5mu]\;\Varid{₂}\;(\Conid{A}\;\mathbin{:}\;\Conid{Set})\;\mathbin{:}\;\Conid{Set}\;\Keyword{where}{}\<[E]%
\\
\>[B]{}\hsindent{3}{}\<[3]%
\>[3]{}\Varid{nil}\;[\mskip1.5mu \mskip1.5mu]\;\mathbin{:}\;\Conid{Expr}\;[\mskip1.5mu \mskip1.5mu]\;\Varid{₂}\;\Conid{A}{}\<[E]%
\\
\>[B]{}\hsindent{3}{}\<[3]%
\>[3]{}\Varid{\char95 !!\char95 }\;\mathbin{:}\;\Conid{Expr}\;[\mskip1.5mu \mskip1.5mu]\;\Varid{₂}\;\Conid{A}{}\<[E]%
\\
\>[B]{}\hsindent{3}{}\<[3]%
\>[3]{}\anonymous \;[\mskip1.5mu \anonymous \mskip1.5mu]\;\Varid{≔\char95 }\;\mathbin{:}\;\Conid{A}\;\Varid{→}\;\Conid{A}\;\Varid{→}\;\Conid{A}\;\Varid{→}\;\Conid{Expr}\;[\mskip1.5mu \mskip1.5mu]\;\Varid{₂}\;\Conid{A}{}\<[E]%
\\
\>[B]{}\Keyword{data}\;\Conid{ExprOption}\;(\Conid{A}\;\mathbin{:}\;\Conid{Set})\;\mathbin{:}\;\Conid{Set}\;\Keyword{where}{}\<[E]%
\\
\>[B]{}\hsindent{3}{}\<[3]%
\>[3]{}\Varid{esome}\;\mathbin{:}\;\Conid{ℕ}\;\Varid{→}\;\Conid{ExprOption}\;\Conid{A}{}\<[E]%
\\
\>[B]{}\hsindent{3}{}\<[3]%
\>[3]{}\Varid{enone}\;\mathbin{:}\;\Conid{ℕ}\;\Varid{→}\;\Conid{ExprOption}\;\Conid{A}{}\<[E]%
\\
\>[B]{}\Keyword{data}\;\Conid{Lit}\;(\Conid{A}\;\mathbin{:}\;\Conid{Set})\;\mathbin{:}\;\Conid{Set}\;\Keyword{where}{}\<[E]%
\\
\>[B]{}\hsindent{3}{}\<[3]%
\>[3]{}\Varid{atom}\;\mathbin{:}\;\Conid{ℕ}\;\Varid{→}\;\Conid{Lit}\;\Conid{A}{}\<[E]%
\ColumnHook
\end{hscode}\resethooks
We then introduce recursion as follows, combining components as a disjoint sum,
written $-\uplus-$ in Agda.
\begin{hscode}\SaveRestoreHook
\column{B}{@{}>{\hspre}l<{\hspost}@{}}%
\column{3}{@{}>{\hspre}l<{\hspost}@{}}%
\column{8}{@{}>{\hspre}l<{\hspost}@{}}%
\column{E}{@{}>{\hspre}l<{\hspost}@{}}%
\>[B]{}\Keyword{data}\;\Conid{RecExpr}\;\mathbin{:}\;\Conid{Set}\;\Keyword{where}{}\<[E]%
\\
\>[B]{}\hsindent{3}{}\<[3]%
\>[3]{}\Varid{expr}\;\mathbin{:}\;\Conid{Lit}\;\Conid{RecExpr}\;{}\<[E]%
\\
\>[3]{}\hsindent{5}{}\<[8]%
\>[8]{}\Varid{⊎}\;\Conid{Expr+₂}\;\Conid{RecExpr}\;{}\<[E]%
\\
\>[3]{}\hsindent{5}{}\<[8]%
\>[8]{}\Varid{⊎}\;\Conid{Expr}\;[\mskip1.5mu \mskip1.5mu]\;\Varid{₂}\;\Conid{RecExpr}\;{}\<[E]%
\\
\>[3]{}\hsindent{5}{}\<[8]%
\>[8]{}\Varid{⊎}\;\Conid{ExprOption}\;\Conid{RecExpr}\;{}\<[E]%
\\
\>[3]{}\hsindent{5}{}\<[8]%
\>[8]{}\Varid{→}\;\Conid{RecExpr}{}\<[E]%
\ColumnHook
\end{hscode}\resethooks
More generally, this type of data can be captured using a ``categorical
approach'' where recursion is introduced as the fixed point of a functor:
% lhs2TeX recognizes > (so will format) but agda doesn't (so won't error).
\begin{hscode}\SaveRestoreHook
\column{B}{@{}>{\hspre}l<{\hspost}@{}}%
\column{3}{@{}>{\hspre}l<{\hspost}@{}}%
\column{4}{@{}>{\hspre}l<{\hspost}@{}}%
\column{9}{@{}>{\hspre}l<{\hspost}@{}}%
\column{E}{@{}>{\hspre}l<{\hspost}@{}}%
\>[3]{}\Keyword{data}\;\Varid{μ\char95 }\;(\Conid{F}\;\mathbin{:}\;\Conid{Set}\;\Varid{→}\;\Conid{Set})\;\mathbin{:}\;\Conid{Set}\;\Keyword{where}{}\<[E]%
\\
\>[3]{}\hsindent{1}{}\<[4]%
\>[4]{}\Varid{inn}\;\mathbin{:}\;\Conid{F}\;(\Varid{μ}\;\Conid{F})\;\Varid{→}\;\Varid{μ}\;\Conid{F}{}\<[E]%
\\
\>[3]{}\Conid{Expr'}\;\mathrel{=}\;\Varid{λ}\;(\Conid{A}\;\mathbin{:}\;\Conid{Set})\;\Varid{→}\;\Conid{Lit}\;\Conid{A}\;\Varid{⊎}\;\Conid{Expr+₂}\;\Conid{A}\;{}\<[E]%
\\
\>[3]{}\hsindent{6}{}\<[9]%
\>[9]{}\Varid{⊎}\;\Conid{Expr}\;[\mskip1.5mu \mskip1.5mu]\;\Varid{₂}\;\Conid{A}\;{}\<[E]%
\\
\>[3]{}\hsindent{6}{}\<[9]%
\>[9]{}\Varid{⊎}\;\Conid{ExprOption}\;\Conid{A}{}\<[E]%
\\
\>[3]{}\Conid{Expr}\;{}\<[9]%
\>[9]{}\mathrel{=}\;\Varid{μ}\;\Conid{Expr'}{}\<[E]%
\ColumnHook
\end{hscode}\resethooks
It is easy to see that this new type is equivalent to {\tt MonolithicExpr} up to
isomorphism
\begin{align*}
Expr &= \mu \, Expr' \\
     &= Expr' \; (\mu \, Expr') \\
     &= Expr' \; Expr \\
     &\cong atom \\
     &\quad \textpipe \; esome \; \textpipe \; enone \\
     &\quad \textpipe \; Expr + Expr \\
     &\quad \textpipe \; nil[] \; \textpipe \; -!!- \; \textpipe \; -[-]:=- \\
\end{align*}

\subsection{Functors and Agda}
The functor $F$, passed into $\mu-$ above, 
serves as the key abstraction allowing us to represent expressions as least
fixed points.
Functors are a special mapping defined over both types and functions satisfying
the so called \emph{functor laws}; a functor F
\begin{enumerate}
  \item assigns to each type $A$, a type $F\,A$
  \item assigns to each function $f : A \rightarrow B$, a function
    $\map f : F A \rightarrow F B$
\end{enumerate}
such that
\begin{enumerate}
  \item identity is preserved: $\map \id = \id$, and
  \item when $f \circ g$ is defined: $\map (f \circ g) = \map f \circ \map g$.
\end{enumerate}
One familiar example is the $List$ functor mapping each type $A$ to
$List\,A$ and each function $f : A \rightarrow B$ to
$\map f : List\,A \rightarrow List\,B$ which applies $f$ to each element of
a list.
Here we define the least fixed point over a restricted class of functors called
the \emph{polynomial functors}. Polynomial functors are a subset roughly
equivalent to the more familiar algebraic polynomials,
\begin{equation*}
  \sum_{n \in N \subseteq \mathbb{N}} A_n X^n
\end{equation*}
where addition is disjoint sum and multiplication is cartesian product.
In Agda, Ulf Norell\cite{dependently typed programming in agda} expresses this
class as a datatype $Functor$ along with an interpretation as a set $[-]$
\begin{hscode}\SaveRestoreHook
\column{B}{@{}>{\hspre}l<{\hspost}@{}}%
\column{3}{@{}>{\hspre}l<{\hspost}@{}}%
\column{E}{@{}>{\hspre}l<{\hspost}@{}}%
\>[B]{}\Keyword{infixl}\;\Varid{6}\;\Varid{\char95 ⊕\char95 }{}\<[E]%
\\
\>[B]{}\Keyword{infixr}\;\Varid{7}\;\Varid{\char95 ⊗\char95 }{}\<[E]%
\\
\>[B]{}\Keyword{data}\;\Conid{Functor}\;\mathbin{:}\;\Conid{Set₁}\;\Keyword{where}{}\<[E]%
\\
\>[B]{}\hsindent{3}{}\<[3]%
\>[3]{}\Conid{X}\;\mathbin{:}\;\Conid{Functor}{}\<[E]%
\\
\>[B]{}\hsindent{3}{}\<[3]%
\>[3]{}\Conid{A}\;\mathbin{:}\;\Conid{Set}\;\Varid{→}\;\Conid{Functor}{}\<[E]%
\\
\>[B]{}\hsindent{3}{}\<[3]%
\>[3]{}\Varid{\char95 ⊕\char95 }\;\mathbin{:}\;\Conid{Functor}\;\Varid{→}\;\Conid{Functor}\;\Varid{→}\;\Conid{Functor}{}\<[E]%
\\
\>[B]{}\hsindent{3}{}\<[3]%
\>[3]{}\Varid{\char95 ⊗\char95 }\;\mathbin{:}\;\Conid{Functor}\;\Varid{→}\;\Conid{Functor}\;\Varid{→}\;\Conid{Functor}{}\<[E]%
\\
\>[B]{}[\mskip1.5mu \anonymous \mskip1.5mu]\;\mathbin{:}\;\Conid{Functor}\;\Varid{→}\;\Conid{Set}\;\Varid{→}\;\Conid{Set}{}\<[E]%
\\
\>[B]{}[\mskip1.5mu \Conid{X}\mskip1.5mu]\;\Conid{B}\;\mathrel{=}\;\Conid{B}{}\<[E]%
\\
\>[B]{}[\mskip1.5mu \Conid{A}\;\Conid{C}\mskip1.5mu]\;\Conid{B}\;\mathrel{=}\;\Conid{C}{}\<[E]%
\\
\>[B]{}[\mskip1.5mu \Conid{F}\;\Varid{⊕}\;\Conid{G}\mskip1.5mu]\;\Conid{B}\;\mathrel{=}\;[\mskip1.5mu \Conid{F}\mskip1.5mu]\;\Conid{B}\;\Varid{⊎}\;[\mskip1.5mu \Conid{G}\mskip1.5mu]\;\Conid{B}{}\<[E]%
\\
\>[B]{}[\mskip1.5mu \Conid{F}\;\Varid{⊗}\;\Conid{G}\mskip1.5mu]\;\Conid{B}\;\mathrel{=}\;[\mskip1.5mu \Conid{F}\mskip1.5mu]\;\Conid{B}\;\Varid{×}\;[\mskip1.5mu \Conid{G}\mskip1.5mu]\;\Conid{B}{}\<[E]%
\ColumnHook
\end{hscode}\resethooks
with least fixed point
\begin{hscode}\SaveRestoreHook
\column{B}{@{}>{\hspre}l<{\hspost}@{}}%
\column{3}{@{}>{\hspre}l<{\hspost}@{}}%
\column{E}{@{}>{\hspre}l<{\hspost}@{}}%
\>[B]{}\Keyword{data}\;\Varid{μ\char95 }\;(\Conid{F}\;\mathbin{:}\;\Conid{Functor})\;\mathbin{:}\;\Conid{Set}\;\Keyword{where}{}\<[E]%
\\
\>[B]{}\hsindent{3}{}\<[3]%
\>[3]{}\Varid{inn}\;\mathbin{:}\;[\mskip1.5mu \Conid{F}\mskip1.5mu]\;(\Varid{μ}\;\Conid{F})\;\Varid{→}\;\Varid{μ}\;\Conid{F}{}\<[E]%
\ColumnHook
\end{hscode}\resethooks
Then to reexpress $Expr$ as a polynomial functor we use sum $-\oplus-$ to define
cases within a type, and product $-\otimes-$ to represent arguments of a
particular case
\begin{hscode}\SaveRestoreHook
\column{B}{@{}>{\hspre}l<{\hspost}@{}}%
\column{E}{@{}>{\hspre}l<{\hspost}@{}}%
\>[B]{}\Conid{Option₁}\;\mathbin{:}\;\Conid{Functor}{}\<[E]%
\\
\>[B]{}\Conid{Option₁}\;\mathrel{=}\;\Conid{X}\;\Varid{⊕}\;\Conid{A}\;\Varid{⊤}{}\<[E]%
\\
\>[B]{}\Conid{Array₁}\;\mathbin{:}\;\Conid{Functor}{}\<[E]%
\\
\>[B]{}\Conid{Array₁}\;\mathrel{=}\;\Conid{X}\;\Varid{⊕}\;\Conid{X}\;\Varid{⊕}\;\Conid{X}\;\Varid{⊕}\;\Conid{A}\;\Varid{⊤}\;\Varid{⊕}\;\Conid{X}\;\Varid{⊕}\;\Conid{X}{}\<[E]%
\\
\>[B]{}\Conid{Sum₁}\;\mathbin{:}\;\Conid{Functor}{}\<[E]%
\\
\>[B]{}\Conid{Sum₁}\;\mathrel{=}\;\Conid{X}\;\Varid{⊗}\;\Conid{X}{}\<[E]%
\\
\>[B]{}\Conid{F₁}\;\mathbin{:}\;\Conid{Functor}{}\<[E]%
\\
\>[B]{}\Conid{F₁}\;\mathrel{=}\;\Conid{A}\;\Conid{ℕ}\;\Varid{⊕}\;\Conid{Option₁}\;\Varid{⊕}\;\Conid{Sum₁}\;\Varid{⊕}\;\Conid{Array₁}{}\<[E]%
\\
\>[B]{}\Conid{E₁}\;\mathbin{:}\;\Conid{Set}{}\<[E]%
\\
\>[B]{}\Conid{E₁}\;\mathrel{=}\;\Varid{μ}\;\Conid{F₁}{}\<[E]%
\ColumnHook
\end{hscode}\resethooks
Unfolding $E_1$ yields the same value calculated above---as we should hope!\begin{hscode}\SaveRestoreHook
\column{B}{@{}>{\hspre}l<{\hspost}@{}}%
\column{3}{@{}>{\hspre}l<{\hspost}@{}}%
\column{5}{@{}>{\hspre}l<{\hspost}@{}}%
\column{E}{@{}>{\hspre}l<{\hspost}@{}}%
\>[3]{}\Conid{E₁}\;\mathrel{=}\;\Varid{μ}\;\Conid{F₁}\;{}\<[E]%
\\
\>[3]{}\hsindent{2}{}\<[5]%
\>[5]{}\mathrel{=}\;[\mskip1.5mu \Conid{A}\;\Conid{ℕ}\;\Varid{⊕}\;\Conid{Option₁}\;\Varid{⊕}\;\Conid{Sum₁}\;\Varid{⊕}\;\Conid{Array₁}\mskip1.5mu]\;{}\<[E]%
\\
\>[3]{}\hsindent{2}{}\<[5]%
\>[5]{}\mathrel{=}\;[\mskip1.5mu \Conid{A}\;\Conid{ℕ}\mskip1.5mu]\;(\Varid{μ}\;\Conid{F₁})\;{}\<[E]%
\\
\>[3]{}\hsindent{2}{}\<[5]%
\>[5]{}\Varid{⊎}\;[\mskip1.5mu \Conid{Option₁}\mskip1.5mu]\;(\Varid{μ}\;\Conid{F₁})\;{}\<[E]%
\\
\>[3]{}\hsindent{2}{}\<[5]%
\>[5]{}\Varid{⊎}\;[\mskip1.5mu \Conid{Sum₁}\mskip1.5mu]\;(\Varid{μ}\;\Conid{F₁})\;{}\<[E]%
\\
\>[3]{}\hsindent{2}{}\<[5]%
\>[5]{}\Varid{⊎}\;[\mskip1.5mu \Conid{Array₁}\mskip1.5mu]\;(\Varid{μ}\;\Conid{F₁})\;{}\<[E]%
\\
\>[3]{}\hsindent{2}{}\<[5]%
\>[5]{}\mathrel{=}\;\Conid{ℕ}\;{}\<[E]%
\\
\>[3]{}\hsindent{2}{}\<[5]%
\>[5]{}\Varid{⊎}\;(\Varid{μ}\;\Conid{F₁})\;\Varid{×}\;\Varid{⊤}\;{}\<[E]%
\\
\>[3]{}\hsindent{2}{}\<[5]%
\>[5]{}\Varid{⊎}\;(\Varid{μ}\;\Conid{F₁})\;\Varid{×}\;(\Varid{μ}\;\Conid{F₁})\;{}\<[E]%
\\
\>[3]{}\hsindent{2}{}\<[5]%
\>[5]{}\Varid{⊎}\;(\Varid{μ}\;\Conid{F₁})\;\Varid{×}\;(\Varid{μ}\;\Conid{F₁})\;\Varid{⊎}\;(\Varid{μ}\;\Conid{F₁})\;\Varid{⊎}\;\Varid{⊤}\;\Varid{⊎}\;(\Varid{μ}\;\Conid{F₁})\;\Varid{⊎}\;(\Varid{μ}\;\Conid{F₁}){}\<[E]%
\ColumnHook
\end{hscode}\resethooks
What do values in $E_1$ look like?  Written directly they appear
nonsensical, consider $6 + 7$
\begin{hscode}\SaveRestoreHook
\column{B}{@{}>{\hspre}l<{\hspost}@{}}%
\column{3}{@{}>{\hspre}l<{\hspost}@{}}%
\column{5}{@{}>{\hspre}l<{\hspost}@{}}%
\column{E}{@{}>{\hspre}l<{\hspost}@{}}%
\>[B]{}\Varid{the-sum}\;\mathbin{:}\;\Conid{E₁}{}\<[E]%
\\
\>[B]{}\Varid{the-sum}\;\mathrel{=}\;\Varid{inn}\;(\Varid{inj₁}\;(\Varid{inj₂}\;({}\<[E]%
\\
\>[B]{}\hsindent{5}{}\<[5]%
\>[5]{}(\Varid{inn}\;(\Varid{inj₁}\;(\Varid{inj₁}\;(\Varid{inj₁}\;\Varid{6})))){}\<[E]%
\\
\>[B]{}\hsindent{3}{}\<[3]%
\>[3]{},(\Varid{inn}\;(\Varid{inj₁}\;(\Varid{inj₁}\;(\Varid{inj₁}\;\Varid{7}))))))){}\<[E]%
\ColumnHook
\end{hscode}\resethooks
Notice here the role that the injections and $inn$ functions play.
Traditionally we would provide a unique name for each branch in an algebraic
datatype, however here we only have two names $inj_1$ and $inj_2$ so we
instead rely on nesting to create unique prefixes. Once we have tagged a
value we must give it a well known type so that parent expressions can
expect a common child type, this is the role of $inn$.  Although cumbersome
we can hide much of this complexity provided the right abstractions
\begin{hscode}\SaveRestoreHook
\column{B}{@{}>{\hspre}l<{\hspost}@{}}%
\column{3}{@{}>{\hspre}l<{\hspost}@{}}%
\column{9}{@{}>{\hspre}l<{\hspost}@{}}%
\column{E}{@{}>{\hspre}l<{\hspost}@{}}%
\>[B]{}\Varid{the-sum'}\;\mathbin{:}\;\Conid{E₁}{}\<[E]%
\\
\>[B]{}\Varid{the-sum'}\;\mathrel{=}\;\Varid{nat₁}\;\Varid{6}\;\Varid{+₁}\;\Varid{nat₁}\;\Varid{7}{}\<[E]%
\\
\>[B]{}\hsindent{3}{}\<[3]%
\>[3]{}\Keyword{where}\;\Varid{nat₁}\;\mathbin{:}\;\Conid{ℕ}\;\Varid{→}\;\Conid{E₁}{}\<[E]%
\\
\>[3]{}\hsindent{6}{}\<[9]%
\>[9]{}\Varid{nat₁}\;\mathrel{=}\;\Varid{inn}\;\Varid{∘}\;\Varid{inj₁}\;\Varid{∘}\;\Varid{inj₁}\;\Varid{∘}\;\Varid{inj₁}{}\<[E]%
\\
\>[3]{}\hsindent{6}{}\<[9]%
\>[9]{}\Varid{\char95 +₁\char95 }\;\mathbin{:}\;\Conid{E₁}\;\Varid{→}\;\Conid{E₁}\;\Varid{→}\;\Conid{E₁}{}\<[E]%
\\
\>[3]{}\hsindent{6}{}\<[9]%
\>[9]{}\Varid{e₁}\;\Varid{+₁}\;\Varid{e₂}\;\mathrel{=}\;\Varid{inn}\;(\Varid{inj₁}\;(\Varid{inj₂}\;(\Varid{e₁},\Varid{e₂}))){}\<[E]%
\ColumnHook
\end{hscode}\resethooks

\section{Syntax and Evaluation Semantics}
We are now ready to define a simple language and its operational semantics.
The language is small including just sums, an option type, and an array with
assignment and lookup.  In Agda, the unit type is written $\top$ and has only
one member: $tt$. $\top$ is used to represent constructors that take no
arguments such as $nil$, the empty list.
\begin{hscode}\SaveRestoreHook
\column{B}{@{}>{\hspre}l<{\hspost}@{}}%
\column{E}{@{}>{\hspre}l<{\hspost}@{}}%
\>[B]{}\Conid{Option}\;\mathbin{:}\;\Conid{Functor}{}\<[E]%
\\
\>[B]{}\Conid{Option}\;\mathrel{=}\;\Conid{X}\;\Varid{⊕}\;\Conid{A}\;\Varid{⊤}{}\<[E]%
\\[\blanklineskip]%
\>[B]{}\Conid{Array}\;\mathbin{:}\;\Conid{Functor}{}\<[E]%
\\
\>[B]{}\Conid{Array}\;\mathrel{=}\;\Conid{X}\;\Varid{⊗}\;\Conid{X}\;\Varid{⊗}\;\Conid{X}\;\Varid{⊕}\;\Conid{A}\;\Varid{⊤}\;\Varid{⊕}\;\Conid{X}\;\Varid{⊗}\;\Conid{X}{}\<[E]%
\\[\blanklineskip]%
\>[B]{}\Conid{Sum}\;\mathbin{:}\;\Conid{Functor}{}\<[E]%
\\
\>[B]{}\Conid{Sum}\;\mathrel{=}\;\Conid{X}\;\Varid{⊗}\;\Conid{X}{}\<[E]%
\\[\blanklineskip]%
\>[B]{}\Conid{FExpr}\;\mathbin{:}\;\Conid{Functor}{}\<[E]%
\\
\>[B]{}\Conid{FExpr}\;\mathrel{=}\;\Conid{A}\;\Conid{ℕ}\;\Varid{⊕}\;\Conid{Option}\;\Varid{⊕}\;\Conid{Sum}\;\Varid{⊕}\;\Conid{Array}{}\<[E]%
\\[\blanklineskip]%
\>[B]{}\Conid{Expr}\;\mathbin{:}\;\Conid{Set}{}\<[E]%
\\
\>[B]{}\Conid{Expr}\;\mathrel{=}\;\Varid{μ}\;\Conid{FExpr}{}\<[E]%
\ColumnHook
\end{hscode}\resethooks
What do each of these definitions mean?  The maybe type has two constructors:
some, which wraps a single expression; and none taking no arguments.  
We define more descriptive constructors for tagging these two types of values
\begin{hscode}\SaveRestoreHook
\column{B}{@{}>{\hspre}l<{\hspost}@{}}%
\column{E}{@{}>{\hspre}l<{\hspost}@{}}%
\>[B]{}\Varid{none₁}\;\mathbin{:}\;\Conid{Expr}{}\<[E]%
\\
\>[B]{}\Varid{none₁}\;\mathrel{=}\;\Varid{inn}\;(\Varid{inj₁}\;(\Varid{inj₁}\;(\Varid{inj₂}\;(\Varid{inj₂}\;\Varid{tt})))){}\<[E]%
\\
\>[B]{}\Varid{some₁}\;\mathbin{:}\;\Conid{Expr}\;\Varid{→}\;\Conid{Expr}{}\<[E]%
\\
\>[B]{}\Varid{some₁}\;\mathrel{=}\;\Varid{inn}\;\Varid{∘}\;\Varid{inj₁}\;\Varid{∘}\;\Varid{inj₁}\;\Varid{∘}\;\Varid{inj₂}\;\Varid{∘}\;\Varid{inj₁}{}\<[E]%
\ColumnHook
\end{hscode}\resethooks
Giving a convenient constructor for $-+-$ is similarly straightforward
\begin{hscode}\SaveRestoreHook
\column{B}{@{}>{\hspre}l<{\hspost}@{}}%
\column{E}{@{}>{\hspre}l<{\hspost}@{}}%
\>[B]{}\Varid{enat}\;\mathbin{:}\;\Conid{ℕ}\;\Varid{→}\;\Conid{Expr}{}\<[E]%
\\
\>[B]{}\Varid{enat}\;\mathrel{=}\;\Varid{inn}\;\Varid{∘}\;\Varid{inj₁}\;\Varid{∘}\;\Varid{inj₁}\;\Varid{∘}\;\Varid{inj₁}{}\<[E]%
\\
\>[B]{}\Varid{\char95 ∔\char95 }\;\mathbin{:}\;\Conid{Expr}\;\Varid{→}\;\Conid{Expr}\;\Varid{→}\;\Conid{Expr}{}\<[E]%
\\
\>[B]{}\Varid{e₁}\;\Varid{∔}\;\Varid{e₂}\;\mathrel{=}\;\Varid{inn}\;(\Varid{inj₁}\;(\Varid{inj₂}\;(\Varid{e₁},\Varid{e₂}))){}\<[E]%
\ColumnHook
\end{hscode}\resethooks
and to define arrays we have assignment taking an array, an index, and a value
to assign at that index; nil, the empty array; and lookup which accepts an array
and an index
\begin{hscode}\SaveRestoreHook
\column{B}{@{}>{\hspre}l<{\hspost}@{}}%
\column{E}{@{}>{\hspre}l<{\hspost}@{}}%
\>[B]{}\anonymous \;[\mskip1.5mu \anonymous \mskip1.5mu]\;\Varid{≔₁\char95 }\;\mathbin{:}\;\Conid{Expr}\;\Varid{→}\;\Conid{Expr}\;\Varid{→}\;\Conid{Expr}\;\Varid{→}\;\Conid{Expr}{}\<[E]%
\\
\>[B]{}\Varid{a}\;[\mskip1.5mu \Varid{i}\mskip1.5mu]\;\Varid{≔₁}\;\Varid{e}\;\mathrel{=}\;\Varid{inn}\;(\Varid{inj₂}\;(\Varid{inj₁}\;(\Varid{inj₁}\;(\Varid{a},\Varid{i},\Varid{e})))){}\<[E]%
\\
\>[B]{}\Varid{nil₁}\;\mathbin{:}\;\Conid{Expr}{}\<[E]%
\\
\>[B]{}\Varid{nil₁}\;\mathrel{=}\;\Varid{inn}\;(\Varid{inj₂}\;(\Varid{inj₁}\;(\Varid{inj₂}\;\Varid{tt}))){}\<[E]%
\\
\>[B]{}\Varid{\char95 !₁\char95 }\;\mathbin{:}\;\Conid{Expr}\;\Varid{→}\;\Conid{Expr}\;\Varid{→}\;\Conid{Expr}{}\<[E]%
\\
\>[B]{}\Varid{a}\;\Varid{!₁}\;\Varid{i}\;\mathrel{=}\;\Varid{inn}\;(\Varid{inj₂}\;(\Varid{inj₂}\;(\Varid{a},\Varid{i}))){}\<[E]%
\ColumnHook
\end{hscode}\resethooks

So far the definition of our syntax has used fairly standard techniques but
we have failed to give any sort of meaning to these expressions.  
We first define a monolithic static and dynamic semantics for this
language, then show how to modularize their definition later
in this section. 
Figure~\ref{fig:value typing rules} defines a simple set of typing rules
using metavariables $e$ to range over experssions and $n$ to range over
values; Figure~\ref{fig:evaluation semantics} gives a small step operational
semantics.
\begin{figure*}[t!]
  \centering
  \begin{tabular}{c c c}
    % Syntax
    \subfloat[Syntax\label{fig:syntax}]{
      $\begin{aligned}
      -∔- &\in Expr \rightarrow Expr \rightarrow Expr \\
      n &\in ℕ \in Expr \\
      -[-]:=- &\in Expr \rightarrow Expr \rightarrow Expr \rightarrow Expr \\
      -!- &\in Expr \rightarrow Expr \rightarrow Expr \\
      nil &\in Expr
      \end{aligned}$
    } &
    \subfloat[Evaluation Semantics\label{fig:evaluation semantics}]{
      \begin{tabular}{c}
        \inference[(stepl+)]
          {e_1 \longrightarrow e_1'}
          {e_1 \dot{+} e_2 \longrightarrow e_1' \dot{+} e_2} \\
        \inference[(stepr+)]
          {e_2 \longrightarrow e_2'}
          {n_1 \dot{+} e_2 \longrightarrow n_1 \dot{+} e_2'} \\
        \inference[(sum)]{}{n_1 \dot{+} n_2 \longrightarrow n_1 + n_2} \\
        \inference[(stepi)]
          {e \longrightarrow e'}
          {a!e \longrightarrow a!e'} \\
        \inference[(lookup)]
          {}
          {a!n \longrightarrow L \llbracket a, n \rrbracket} \\
      \end{tabular}
    } \\%
    
    \multicolumn{2}{c}{
      \rule{\columnwidth}{1px}
    } \\%
    
    % Values
    \subfloat[Value Typing\label{fig:value typing rules}]{
      \inference[(ok-value)]{}{n : Nat}
    } &
    % Sums
    \subfloat[Sum Typing\label{fig:sum typing rules}]{
      \begin{tabular}{c}
        \inference[(ok-sum)]
          {e_1 : Nat & e_2 : Nat}
          {e_1 \dot{+} e_2 : Nat} \\
      \end{tabular}
    } \\
    % Arrays
    \multicolumn{2}{c}{
      \subfloat[Array Typing\label{fig:sum typing rules}]{
        \begin{tabular}{c}
          \inference[(ok-nil)]{}{nil : Array} \\
          \inference[(ok-lookup)]
            {a : Array & e : Nat}
            {a!e : Option} \\
          \inference[(ok-ins)]
            {a : Array & n : Nat & e : Nat}
            {a[n] := e : Array}
        \end{tabular}
      }
    }
  \end{tabular}
  \label{fig:language definition}
\end{figure*}

While Agda is expressive enough to implement these rules, directly and
indeed they are nearly a direct reflection of that implementation,
recall that our goal is to create several independant languages each
carrying their own semantics.  We begin by defining monolithic
semantics for $Expr$ and proceed to determine points of failure and to
dissect the definition into independant constituents. To simplify
things we define our notion of $Type$ as a closed ADT
\begin{hscode}\SaveRestoreHook
\column{B}{@{}>{\hspre}l<{\hspost}@{}}%
\column{3}{@{}>{\hspre}l<{\hspost}@{}}%
\column{E}{@{}>{\hspre}l<{\hspost}@{}}%
\>[B]{}\Keyword{data}\;\Conid{Type}\;\mathbin{:}\;\Conid{Set}\;\Keyword{where}{}\<[E]%
\\
\>[B]{}\hsindent{3}{}\<[3]%
\>[3]{}\Conid{TArray}\;\mathbin{:}\;\Conid{Type}{}\<[E]%
\\
\>[B]{}\hsindent{3}{}\<[3]%
\>[3]{}\Conid{TOption}\;\mathbin{:}\;\Conid{Type}{}\<[E]%
\\
\>[B]{}\hsindent{3}{}\<[3]%
\>[3]{}\Conid{TNat}\;\mathbin{:}\;\Conid{Type}{}\<[E]%
\ColumnHook
\end{hscode}\resethooks
and here is the definition of the monolithic type system
and evaluation relation in Agda.
\begin{hscode}\SaveRestoreHook
\column{B}{@{}>{\hspre}l<{\hspost}@{}}%
\column{3}{@{}>{\hspre}l<{\hspost}@{}}%
\column{5}{@{}>{\hspre}l<{\hspost}@{}}%
\column{E}{@{}>{\hspre}l<{\hspost}@{}}%
\>[B]{}\Keyword{data}\;\Conid{Welltyped}\;\mathbin{:}\;\Conid{Expr}\;\Varid{→}\;\Conid{Type}\;\Varid{→}\;\Conid{Set₁}\;\Keyword{where}{}\<[E]%
\\
\>[B]{}\hsindent{3}{}\<[3]%
\>[3]{}\Varid{ok-value}\;\mathbin{:}\;\{\mskip1.5mu \Varid{n}\;\mathbin{:}\;\Conid{ℕ}\mskip1.5mu\}\;\Varid{→}\;\Conid{Welltyped}\;(\Varid{enat}\;\Varid{n})\;\Conid{TNat}{}\<[E]%
\\
\>[B]{}\hsindent{3}{}\<[3]%
\>[3]{}\Varid{ok-sum}\;\mathbin{:}\;\{\mskip1.5mu \Varid{e₁}\;\Varid{e₂}\;\mathbin{:}\;\Conid{Expr}\mskip1.5mu\}\;{}\<[E]%
\\
\>[3]{}\hsindent{2}{}\<[5]%
\>[5]{}\Varid{→}\;\Conid{Welltyped}\;\Varid{e₁}\;\Conid{TNat}\;\Varid{→}\;\Conid{Welltyped}\;\Varid{e₂}\;\Conid{TNat}\;{}\<[E]%
\\
\>[3]{}\hsindent{2}{}\<[5]%
\>[5]{}\Varid{→}\;\Conid{Welltyped}\;(\Varid{e₁}\;\Varid{∔}\;\Varid{e₂})\;\Conid{TNat}{}\<[E]%
\\
\>[B]{}\hsindent{3}{}\<[3]%
\>[3]{}\Varid{ok-nil}\;\mathbin{:}\;\Conid{Welltyped}\;\Varid{nil₁}\;\Conid{TArray}{}\<[E]%
\\
\>[B]{}\hsindent{3}{}\<[3]%
\>[3]{}\Varid{ok-lookup}\;\mathbin{:}\;\{\mskip1.5mu \Varid{a}\;\Varid{e}\;\mathbin{:}\;\Conid{Expr}\mskip1.5mu\}\;{}\<[E]%
\\
\>[3]{}\hsindent{2}{}\<[5]%
\>[5]{}\Varid{→}\;\Conid{Welltyped}\;\Varid{a}\;\Conid{TArray}\;{}\<[E]%
\\
\>[3]{}\hsindent{2}{}\<[5]%
\>[5]{}\Varid{→}\;\Conid{Welltyped}\;\Varid{e}\;\Conid{TNat}\;{}\<[E]%
\\
\>[3]{}\hsindent{2}{}\<[5]%
\>[5]{}\Varid{→}\;\Conid{Welltyped}\;(\Varid{a}\;\Varid{!₁}\;\Varid{e})\;\Conid{TOption}{}\<[E]%
\\
\>[B]{}\hsindent{3}{}\<[3]%
\>[3]{}\Varid{ok-ins}\;\mathbin{:}\;\{\mskip1.5mu \Varid{a}\;\Varid{e}\;\Varid{n}\;\mathbin{:}\;\Conid{Expr}\mskip1.5mu\}\;{}\<[E]%
\\
\>[3]{}\hsindent{2}{}\<[5]%
\>[5]{}\Varid{→}\;\Conid{Welltyped}\;\Varid{a}\;\Conid{TArray}\;{}\<[E]%
\\
\>[3]{}\hsindent{2}{}\<[5]%
\>[5]{}\Varid{→}\;\Conid{Welltyped}\;\Varid{e}\;\Conid{TNat}\;{}\<[E]%
\\
\>[3]{}\hsindent{2}{}\<[5]%
\>[5]{}\Varid{→}\;\Conid{Welltyped}\;\Varid{n}\;\Conid{TNat}\;{}\<[E]%
\\
\>[3]{}\hsindent{2}{}\<[5]%
\>[5]{}\Varid{→}\;\Conid{Welltyped}\;(\Varid{a}\;[\mskip1.5mu \Varid{n}\mskip1.5mu]\;\Varid{≔₁}\;\Varid{e})\;\Conid{TArray}{}\<[E]%
\\[\blanklineskip]%
\>[B]{}\Keyword{infix}\;\Varid{2}\;\Varid{\char95 ⟶E\char95 }{}\<[E]%
\\
\>[B]{}\Keyword{data}\;\Varid{\char95 ⟶E\char95 }\;\mathbin{:}\;\Conid{Expr}\;\Varid{→}\;\Conid{Expr}\;\Varid{→}\;\Conid{Set}\;\Keyword{where}{}\<[E]%
\\
\>[B]{}\hsindent{3}{}\<[3]%
\>[3]{}\Varid{stepl}\;\mathbin{:}\;\{\mskip1.5mu \Varid{e₁}\;\Varid{e₁'}\;\Varid{e₂}\;\mathbin{:}\;\Conid{Expr}\mskip1.5mu\}\;{}\<[E]%
\\
\>[3]{}\hsindent{2}{}\<[5]%
\>[5]{}\Varid{→}\;\Varid{e₁}\;\Varid{⟶E}\;\Varid{e₁'}\;{}\<[E]%
\\
\>[3]{}\hsindent{2}{}\<[5]%
\>[5]{}\Varid{→}\;\Varid{e₁}\;\Varid{∔}\;\Varid{e₂}\;\Varid{⟶E}\;\Varid{e₁'}\;\Varid{∔}\;\Varid{e₂}{}\<[E]%
\\
\>[B]{}\hsindent{3}{}\<[3]%
\>[3]{}\Varid{stepr}\;\mathbin{:}\;\{\mskip1.5mu \Varid{n₁}\;\mathbin{:}\;\Conid{ℕ}\mskip1.5mu\}\;\{\mskip1.5mu \Varid{e₂}\;\Varid{e₂'}\;\mathbin{:}\;\Conid{Expr}\mskip1.5mu\}\;{}\<[E]%
\\
\>[3]{}\hsindent{2}{}\<[5]%
\>[5]{}\Varid{→}\;\Varid{e₂}\;\Varid{⟶E}\;\Varid{e₂'}\;{}\<[E]%
\\
\>[3]{}\hsindent{2}{}\<[5]%
\>[5]{}\Varid{→}\;\Varid{enat}\;\Varid{n₁}\;\Varid{∔}\;\Varid{e₂}\;\Varid{⟶E}\;\Varid{enat}\;\Varid{n₁}\;\Varid{∔}\;\Varid{e₂'}{}\<[E]%
\\
\>[B]{}\hsindent{3}{}\<[3]%
\>[3]{}\Varid{sum}\;\mathbin{:}\;\{\mskip1.5mu \Varid{n₁}\;\Varid{n₂}\;\mathbin{:}\;\Conid{ℕ}\mskip1.5mu\}\;{}\<[E]%
\\
\>[3]{}\hsindent{2}{}\<[5]%
\>[5]{}\Varid{→}\;\Varid{enat}\;\Varid{n₁}\;\Varid{∔}\;\Varid{enat}\;\Varid{n₂}\;\Varid{⟶E}\;\Varid{enat}\;(\Varid{n₁}\;\Varid{+ℕ}\;\Varid{n₂}){}\<[E]%
\\
\>[B]{}\hsindent{3}{}\<[3]%
\>[3]{}\Varid{stepi}\;\mathbin{:}\;\{\mskip1.5mu \Varid{e}\;\Varid{e'}\;\Varid{a}\;\mathbin{:}\;\Conid{Expr}\mskip1.5mu\}\;{}\<[E]%
\\
\>[3]{}\hsindent{2}{}\<[5]%
\>[5]{}\Varid{→}\;\Varid{e}\;\Varid{⟶E}\;\Varid{e'}\;{}\<[E]%
\\
\>[3]{}\hsindent{2}{}\<[5]%
\>[5]{}\Varid{→}\;\Varid{a}\;\Varid{!₁}\;\Varid{e}\;\Varid{⟶E}\;\Varid{a}\;\Varid{!₁}\;\Varid{e'}{}\<[E]%
\\
\>[B]{}\hsindent{3}{}\<[3]%
\>[3]{}\Varid{lookup}\;\mathbin{:}\;\{\mskip1.5mu \Varid{a}\;\Varid{n}\;\mathbin{:}\;\Conid{Expr}\mskip1.5mu\}\;{}\<[E]%
\\
\>[3]{}\hsindent{2}{}\<[5]%
\>[5]{}\Varid{→}\;\Varid{a}\;\Varid{!₁}\;\Varid{n}\;\Varid{⟶E}\;\Conid{L⟦}\;\Varid{a},\Varid{n}\;\Varid{⟧₁}{}\<[E]%
\ColumnHook
\end{hscode}\resethooks
The function $L\llbracket -,- \rrbracket_1$ is the lookup function
that evaluates to $\some\, a_n$ when $a_n$ has been defined
and $none$ otherwise.
Notice that we currently do not restrict the values of $n$ enough in the
$ok \mhyphen ins$ rule; our typing rules require that n be a value while in
Agda we have only required it be an expression.  Some notion of value is
needed and a common solution is to add a tag $\mathit{Value}$ to the
$\mathit{Expr}$ type and pattern match; here $\mathit{Value}$ is called $[A \mathbb{N}]$ and
in a dependantly typed context we might then define a predicate over $\mathit{Value}$.
However because the sum type has only one type of value, a number, it is simpler
to use $enat$ directly.

This method for defining semantics is common with the advantage of being
direct and concise, but similar to our first implementation of $\mathit{Expr+}$ and
$\mathit{MonolithicExpr}$ above: there is no simple mechanism for code reuse.  The
answer is again to delay recursion.

\subsection{Dissecting the Step Relation}
In order to modularize the evaluation rules we define a separate step relation
for each functor making up our $Expr$ type.  First note that $-\dot{+}-$ doesn't
make use of \emph{how} the step from $e_1$ to $e_2$ occurs so we can factor this
top-level relation\begin{hscode}\SaveRestoreHook
\column{B}{@{}>{\hspre}l<{\hspost}@{}}%
\column{3}{@{}>{\hspre}l<{\hspost}@{}}%
\column{5}{@{}>{\hspre}l<{\hspost}@{}}%
\column{7}{@{}>{\hspre}l<{\hspost}@{}}%
\column{E}{@{}>{\hspre}l<{\hspost}@{}}%
\>[3]{}\Keyword{data}\;\Varid{\char95 ⟶⁺\char95 }\;\{\mskip1.5mu \Varid{\char95 ⟶\char95 }\;\mathbin{:}\;\Conid{Expr}\;\Varid{→}\;\Conid{Expr}\;\Varid{→}\;\Conid{Set}\mskip1.5mu\}{}\<[E]%
\\
\>[3]{}\hsindent{2}{}\<[5]%
\>[5]{}\mathbin{:}\;\Conid{Expr}\;\Varid{→}\;\Conid{Expr}\;\Varid{→}\;\Conid{Set}\;\Keyword{where}{}\<[E]%
\\
\>[3]{}\hsindent{2}{}\<[5]%
\>[5]{}\Varid{stepl}\;\mathbin{:}\;\{\mskip1.5mu \Varid{e₁}\;\Varid{e₁'}\;\Varid{e₂}\;\mathbin{:}\;\Conid{Expr}\mskip1.5mu\}\;{}\<[E]%
\\
\>[5]{}\hsindent{2}{}\<[7]%
\>[7]{}\Varid{→}\;\Varid{e₁}\;\Varid{⟶}\;\Varid{e₁'}\;\Varid{→}\;\Varid{e₁}\;\Varid{∔}\;\Varid{e₂}\;\Varid{⟶⁺}\;\Varid{e₁'}\;\Varid{∔}\;\Varid{e₂}{}\<[E]%
\\
\>[3]{}\hsindent{2}{}\<[5]%
\>[5]{}\Varid{stepr}\;\mathbin{:}\;\{\mskip1.5mu \Varid{n₁}\;\mathbin{:}\;\Conid{ℕ}\mskip1.5mu\}\;\{\mskip1.5mu \Varid{e₂}\;\Varid{e₂'}\;\mathbin{:}\;\Conid{Expr}\mskip1.5mu\}\;{}\<[E]%
\\
\>[5]{}\hsindent{2}{}\<[7]%
\>[7]{}\Varid{→}\;\Varid{e₂}\;\Varid{⟶}\;\Varid{e₂'}\;{}\<[E]%
\\
\>[5]{}\hsindent{2}{}\<[7]%
\>[7]{}\Varid{→}\;\Varid{enat}\;\Varid{n₁}\;\Varid{∔}\;\Varid{e₂}\;\Varid{⟶⁺}\;\Varid{enat}\;\Varid{n₁}\;\Varid{∔}\;\Varid{e₂'}{}\<[E]%
\\
\>[3]{}\hsindent{2}{}\<[5]%
\>[5]{}\Varid{sum}\;\mathbin{:}\;\{\mskip1.5mu \Varid{n₁}\;\Varid{n₂}\;\mathbin{:}\;\Conid{ℕ}\mskip1.5mu\}\;{}\<[E]%
\\
\>[5]{}\hsindent{2}{}\<[7]%
\>[7]{}\Varid{→}\;\Varid{enat}\;\Varid{n₁}\;\Varid{∔}\;\Varid{enat}\;\Varid{n₂}\;\Varid{⟶⁺}\;\Varid{enat}\;(\Varid{n₁}\;\Varid{+ℕ}\;\Varid{n₂}){}\<[E]%
\ColumnHook
\end{hscode}\resethooks
While this is better there is still an undesirable reference to the datatype
$Expr$.  Applying the same factorization here to the underlying functor requires
parametrization by two extra coercion functions, these are the $-\dot{+}-$
and $enat$ functions defined previously.  The new names $lift^{+}$ and
$lift\mathbb{N}$ used here are meant to imply that a subtype is being ``lifted''
into its supertype\begin{hscode}\SaveRestoreHook
\column{B}{@{}>{\hspre}l<{\hspost}@{}}%
\column{3}{@{}>{\hspre}l<{\hspost}@{}}%
\column{5}{@{}>{\hspre}l<{\hspost}@{}}%
\column{7}{@{}>{\hspre}l<{\hspost}@{}}%
\column{E}{@{}>{\hspre}l<{\hspost}@{}}%
\>[3]{}\Keyword{data}\;\Varid{\char95 ⟶⁺\char95 }\;\{\mskip1.5mu \Conid{E}\;\mathbin{:}\;\Conid{Functor}\mskip1.5mu\}\;\{\mskip1.5mu \Varid{\char95 ⟶\char95 }\;\mathbin{:}\;\Varid{μ}\;\Conid{E}\;\Varid{→}\;\Varid{μ}\;\Conid{E}\;\Varid{→}\;\Conid{Set}\mskip1.5mu\}{}\<[E]%
\\
\>[3]{}\hsindent{2}{}\<[5]%
\>[5]{}\{\mskip1.5mu \Varid{lift⁺}\;\mathbin{:}\;[\mskip1.5mu \Conid{Sum}\mskip1.5mu]\;(\Varid{μ}\;\Conid{E})\;\Varid{→}\;\Varid{μ}\;\Conid{E}\mskip1.5mu\}\;\{\mskip1.5mu \Varid{liftℕ}\;\mathbin{:}\;\Conid{ℕ}\;\Varid{→}\;\Varid{μ}\;\Conid{E}\mskip1.5mu\}{}\<[E]%
\\
\>[3]{}\hsindent{2}{}\<[5]%
\>[5]{}\mathbin{:}\;\Varid{μ}\;\Conid{E}\;\Varid{→}\;\Varid{μ}\;\Conid{E}\;\Varid{→}\;\Conid{Set}\;\Keyword{where}{}\<[E]%
\\
\>[3]{}\hsindent{2}{}\<[5]%
\>[5]{}\Varid{stepl}\;\mathbin{:}\;\{\mskip1.5mu \Varid{e₁}\;\Varid{e₁'}\;\Varid{e₂}\;\mathbin{:}\;\Varid{μ}\;\Conid{E}\mskip1.5mu\}\;{}\<[E]%
\\
\>[5]{}\hsindent{2}{}\<[7]%
\>[7]{}\Varid{→}\;\Varid{e₁}\;\Varid{⟶}\;\Varid{e₁'}\;\Varid{→}\;\Varid{lift⁺}\;(\Varid{e₁},\Varid{e₂})\;\Varid{⟶⁺}\;\Varid{lift⁺}\;(\Varid{e₁'},\Varid{e₂}){}\<[E]%
\\
\>[3]{}\hsindent{2}{}\<[5]%
\>[5]{}\Varid{stepr}\;\mathbin{:}\;\{\mskip1.5mu \Varid{n₁}\;\mathbin{:}\;\Conid{ℕ}\mskip1.5mu\}\;\{\mskip1.5mu \Varid{e₂}\;\Varid{e₂'}\;\mathbin{:}\;\Varid{μ}\;\Conid{E}\mskip1.5mu\}\;{}\<[E]%
\\
\>[5]{}\hsindent{2}{}\<[7]%
\>[7]{}\Varid{→}\;\Varid{e₂}\;\Varid{⟶}\;\Varid{e₂'}\;{}\<[E]%
\\
\>[5]{}\hsindent{2}{}\<[7]%
\>[7]{}\Varid{→}\;\Varid{lift⁺}\;(\Varid{liftℕ}\;\Varid{n₁},\Varid{e₂})\;\Varid{⟶⁺}\;\Varid{lift⁺}\;(\Varid{liftℕ}\;\Varid{n₁},\Varid{e₂'}){}\<[E]%
\\
\>[3]{}\hsindent{2}{}\<[5]%
\>[5]{}\Varid{sum}\;\mathbin{:}\;\{\mskip1.5mu \Varid{n₁}\;\Varid{n₂}\;\mathbin{:}\;\Conid{ℕ}\mskip1.5mu\}\;{}\<[E]%
\\
\>[5]{}\hsindent{2}{}\<[7]%
\>[7]{}\Varid{→}\;\Varid{lift⁺}\;(\Varid{liftℕ}\;\Varid{n₁},\Varid{liftℕ}\;\Varid{n₂})\;\Varid{⟶⁺}\;\Varid{liftℕ}\;(\Varid{n₁}\;\Varid{+ℕ}\;\Varid{n₂}){}\<[E]%
\ColumnHook
\end{hscode}\resethooks
Unfortunately this definition falls short too.  When we lift terms into
the expression type $\mu E$, Agda ``forgets'' the constituents $e_1$ and
$e_2$---in turn we lose the ability to reason about these distinct components of
the sums $e_n$ and $e_n'$. This later becomes a problem when, for example,
attempting to abstract the welltyping relation.

An intelligent human can peel away $lift^{+}$ and see that the terms
$e_1$ and $e_2$ in $-\longrightarrow-$ and $-\longrightarrow^{+}-$ are the same
because $lift^{+}$ is \emph{injective}.  However Agda is unconvinced,
and rightfully so, for it does not require a particularly great deal of
ingenuity to find a counterexample, consider taking $E = \mathit{FExpr}$ so that
$\mu E = \mathit{Expr}$\begin{hscode}\SaveRestoreHook
\column{B}{@{}>{\hspre}l<{\hspost}@{}}%
\column{3}{@{}>{\hspre}l<{\hspost}@{}}%
\column{E}{@{}>{\hspre}l<{\hspost}@{}}%
\>[3]{}\Varid{forgetful-lift⁺}\;\mathbin{:}\;[\mskip1.5mu \Conid{Sum}\mskip1.5mu]\;\Conid{Expr}\;\Varid{→}\;\Conid{Expr}{}\<[E]%
\\
\>[3]{}\Varid{forgetful-lift⁺}\;(\Varid{e₁},\Varid{e₂})\;\mathrel{=}\;\Varid{enat}\;\Varid{0}{}\<[E]%
\ColumnHook
\end{hscode}\resethooks
The problem is that  our abstraction is too general.
What we require is a proof that $[Sum](\mu E)$ and $\mathbb{N}$ are
subtypes of the top-level expression datatype $\mu E$.
The solution to the problem is drawn from the notion of a categorical subobject.

We proceed by delaying application of injections and view the objects
as injectable, existential terms.  The importance of this approach is
two-fold: firstly this allows us to take inverses of lift functions while we are
secondly able to retain the perspective of operating on a single type $\mu E$.

\section{Lazy Coercions}
A subobject of a type $T$ is a left invertible function with codomain $T$,
$\mathit{lift} : S \hookrightarrow T$.
Being restricted to polynomial functors, we know that all our subobjects
$\mathit{lift} : S \rightarrow \mu E$ will be some composition of $inn$, $inj_1$
and $inj_2$ so a proof that $S$ is a subtype of $\mu E$ is merely a
description of which direction to move at each point in a disjoin sum
\begin{hscode}\SaveRestoreHook
\column{B}{@{}>{\hspre}l<{\hspost}@{}}%
\column{3}{@{}>{\hspre}l<{\hspost}@{}}%
\column{5}{@{}>{\hspre}l<{\hspost}@{}}%
\column{9}{@{}>{\hspre}l<{\hspost}@{}}%
\column{E}{@{}>{\hspre}l<{\hspost}@{}}%
\>[B]{}\Keyword{infix}\;\Varid{3}\;\Varid{\char95 Contains\char95 }{}\<[E]%
\\
\>[B]{}\Keyword{data}\;\Varid{\char95 Contains\char95 }\;\mathbin{:}\;\Conid{Functor}\;\Varid{→}\;\Conid{Functor}\;\Varid{→}\;\Conid{Set₁}\;\Keyword{where}{}\<[E]%
\\
\>[B]{}\hsindent{3}{}\<[3]%
\>[3]{}\Varid{refl}\;\mathbin{:}\;\{\mskip1.5mu \Conid{F}\;\mathbin{:}\;\Conid{Functor}\mskip1.5mu\}\;{}\<[E]%
\\
\>[3]{}\hsindent{2}{}\<[5]%
\>[5]{}\Varid{→}\;\Conid{F}\;\Conid{Contains}\;\Conid{F}{}\<[E]%
\\
\>[B]{}\hsindent{3}{}\<[3]%
\>[3]{}\Varid{left}\;{}\<[9]%
\>[9]{}\mathbin{:}\;\{\mskip1.5mu \Conid{A}\;\Conid{B}\;\Conid{F}\;\mathbin{:}\;\Conid{Functor}\mskip1.5mu\}\;{}\<[E]%
\\
\>[3]{}\hsindent{2}{}\<[5]%
\>[5]{}\Varid{→}\;\Conid{F}\;\Conid{Contains}\;\Conid{A}\;\Varid{⊕}\;\Conid{B}\;\Varid{→}\;\Conid{F}\;\Conid{Contains}\;\Conid{A}{}\<[E]%
\\
\>[B]{}\hsindent{3}{}\<[3]%
\>[3]{}\Varid{right}\;\mathbin{:}\;\{\mskip1.5mu \Conid{A}\;\Conid{B}\;\Conid{F}\;\mathbin{:}\;\Conid{Functor}\mskip1.5mu\}\;{}\<[E]%
\\
\>[3]{}\hsindent{2}{}\<[5]%
\>[5]{}\Varid{→}\;\Conid{F}\;\Conid{Contains}\;\Conid{A}\;\Varid{⊕}\;\Conid{B}\;\Varid{→}\;\Conid{F}\;\Conid{Contains}\;\Conid{B}{}\<[E]%
\ColumnHook
\end{hscode}\resethooks
Now we can define containment on a functor's interpretation as a set
\begin{hscode}\SaveRestoreHook
\column{B}{@{}>{\hspre}l<{\hspost}@{}}%
\column{3}{@{}>{\hspre}l<{\hspost}@{}}%
\column{5}{@{}>{\hspre}l<{\hspost}@{}}%
\column{E}{@{}>{\hspre}l<{\hspost}@{}}%
\>[B]{}\Keyword{infix}\;\Varid{3}\;\Varid{\char95 ↣\char95 }{}\<[E]%
\\
\>[B]{}\Keyword{data}\;\Varid{\char95 ↣\char95 }\;\mathbin{:}\;\Conid{Set}\;\Varid{→}\;\Conid{Set}\;\Varid{→}\;\Conid{Set₁}\;\Keyword{where}{}\<[E]%
\\
\>[B]{}\hsindent{3}{}\<[3]%
\>[3]{}\Varid{inj}\;\mathbin{:}\;\{\mskip1.5mu \Conid{F}\;\Conid{A}\;\mathbin{:}\;\Conid{Functor}\mskip1.5mu\}\;{}\<[E]%
\\
\>[3]{}\hsindent{2}{}\<[5]%
\>[5]{}\Varid{→}\;\Conid{F}\;\Conid{Contains}\;\Conid{A}\;\Varid{→}\;[\mskip1.5mu \Conid{A}\mskip1.5mu]\;(\Varid{μ}\;\Conid{F})\;\Varid{↣}\;(\Varid{μ}\;\Conid{F}){}\<[E]%
\ColumnHook
\end{hscode}\resethooks
with conversion functions defined as
\begin{hscode}\SaveRestoreHook
\column{B}{@{}>{\hspre}l<{\hspost}@{}}%
\column{E}{@{}>{\hspre}l<{\hspost}@{}}%
\>[B]{}\Varid{upcast}\;\mathbin{:}\;\Varid{∀}\;\{\mskip1.5mu \Conid{F}\;\Conid{A}\mskip1.5mu\}\;\Varid{→}\;\Conid{F}\;\Conid{Contains}\;\Conid{A}\;\Varid{→}\;[\mskip1.5mu \Conid{A}\mskip1.5mu]\;(\Varid{μ}\;\Conid{F})\;\Varid{→}\;\Varid{μ}\;\Conid{F}{}\<[E]%
\\
\>[B]{}\Varid{upcast}\;\Varid{refl}\;\mathrel{=}\;\Varid{inn}{}\<[E]%
\\
\>[B]{}\Varid{upcast}\;(\Varid{left}\;\Varid{t})\;\mathrel{=}\;\Varid{upcast}\;\Varid{t}\;\Varid{∘}\;\Varid{inj₁}{}\<[E]%
\\
\>[B]{}\Varid{upcast}\;(\Varid{right}\;\Varid{t})\;\mathrel{=}\;\Varid{upcast}\;\Varid{t}\;\Varid{∘}\;\Varid{inj₂}{}\<[E]%
\\[\blanklineskip]%
\>[B]{}\Varid{apply}\;\mathbin{:}\;\{\mskip1.5mu \Conid{A}\;\Conid{B}\;\mathbin{:}\;\Conid{Set}\mskip1.5mu\}\;\Varid{→}\;(\Conid{A}\;\Varid{↣}\;\Conid{B})\;\Varid{→}\;\Conid{A}\;\Varid{→}\;\Conid{B}{}\<[E]%
\\
\>[B]{}\Varid{apply}\;(\Varid{inj}\;\Varid{t})\;\mathrel{=}\;\Varid{upcast}\;\Varid{t}{}\<[E]%
\ColumnHook
\end{hscode}\resethooks
Recall the two goals we had in mind.  We first wished to take the inverse of
a lift function to gain access to its arguments, in the case of $-+-$ these
were $e_1$ and $e_2$.  By representing an injection as a delayed application
of a subobject---because the constructor's arguments are stored as a part of the
coercion---finding left inverses will become a trivial case of pattern
matching. To delay function application allowing Agda to effectively peel away
the $lift$ functions we define a $\mathit{LazyCoercion}$ datatype from type $A$ to $B$
representing the \emph{intention} of coercing an object $a \in A$ while
treating it at the type-level as $B$.
A lazy coercion is then an injection
$A \rightarrowtail B$ along with an object in $A$
\begin{hscode}\SaveRestoreHook
\column{B}{@{}>{\hspre}l<{\hspost}@{}}%
\column{3}{@{}>{\hspre}l<{\hspost}@{}}%
\column{E}{@{}>{\hspre}l<{\hspost}@{}}%
\>[B]{}\Keyword{data}\;\Conid{LazyCoercion}\;\mathbin{:}\;\Conid{Set}\;\Varid{→}\;\Conid{Set₁}\;\Keyword{where}{}\<[E]%
\\
\>[B]{}\hsindent{3}{}\<[3]%
\>[3]{}\Varid{inj}\;\mathbin{:}\;\{\mskip1.5mu \Conid{A}\;\Conid{B}\;\mathbin{:}\;\Conid{Set}\mskip1.5mu\}\;\Varid{→}\;(\Conid{A}\;\Varid{↣}\;\Conid{B})\;\Varid{→}\;\Conid{A}\;\Varid{→}\;\Conid{LazyCoercion}\;\Conid{B}{}\<[E]%
\\
\>[B]{}\Varid{coerce}\;\mathbin{:}\;\{\mskip1.5mu \Conid{B}\;\mathbin{:}\;\Conid{Set}\mskip1.5mu\}\;\Varid{→}\;\Conid{LazyCoercion}\;\Conid{B}\;\Varid{→}\;\Conid{B}{}\<[E]%
\\
\>[B]{}\Varid{coerce}\;(\Varid{inj}\;\Varid{f}\;\Varid{e})\;\mathrel{=}\;\Varid{apply}\;\Varid{f}\;\Varid{e}{}\<[E]%
\ColumnHook
\end{hscode}\resethooks
Our second goal was to operate on objects of a single type.
Why is this the case?  Recall that the type of our step relation is indexed
by two expressions: $(e_1 : \mathit{Expr}) \longrightarrow_E (e_2 : \mathit{Expr})$.
We should expect the same of the final abstraction over step relations
because it cannot easily name the underlying type of its indexing expressions.  
Instead we have packaged the indices as \emph{existentials} which are
viewed as the type $B$.

We seem to be close to a modular step relation $-\longrightarrow^{+}-$,
defining at each point another level of abstraction to delay immediate
application.  To modularize datatypes, recursion is delayed and types are viewed
as polynomial functors, then to modularize step relations, evaluation is
parametrized and expression upcasts are delayed by viewing them as an intention.

\section{Defining a Modular Step Relation}
Attempting again to define a step relation for addition we find very little
has changed
\begin{hscode}\SaveRestoreHook
\column{B}{@{}>{\hspre}l<{\hspost}@{}}%
\column{3}{@{}>{\hspre}l<{\hspost}@{}}%
\column{5}{@{}>{\hspre}l<{\hspost}@{}}%
\column{7}{@{}>{\hspre}l<{\hspost}@{}}%
\column{E}{@{}>{\hspre}l<{\hspost}@{}}%
\>[B]{}\Keyword{data}\;\Varid{\char95 ⟶⁺\char95 }\;\{\mskip1.5mu \Conid{E}\;\mathbin{:}\;\Conid{Functor}\mskip1.5mu\}{}\<[E]%
\\
\>[B]{}\hsindent{3}{}\<[3]%
\>[3]{}\{\mskip1.5mu \Varid{\char95 ⟶\char95 }\;\mathbin{:}\;\Varid{μ}\;\Conid{E}\;\Varid{→}\;\Varid{μ}\;\Conid{E}\;\Varid{→}\;\Conid{Set₁}\mskip1.5mu\}{}\<[E]%
\\
\>[B]{}\hsindent{3}{}\<[3]%
\>[3]{}\{\mskip1.5mu \Varid{lift⁺}\;\mathbin{:}\;[\mskip1.5mu \Conid{Sum}\mskip1.5mu]\;(\Varid{μ}\;\Conid{E})\;\Varid{↣}\;\Varid{μ}\;\Conid{E}\mskip1.5mu\}{}\<[E]%
\\
\>[B]{}\hsindent{3}{}\<[3]%
\>[3]{}\{\mskip1.5mu \Varid{liftℕ}\;\mathbin{:}\;\Conid{ℕ}\;\Varid{↣}\;\Varid{μ}\;\Conid{E}\mskip1.5mu\}{}\<[E]%
\\
\>[B]{}\hsindent{3}{}\<[3]%
\>[3]{}\mathbin{:}\;\Conid{LazyCoercion}\;(\Varid{μ}\;\Conid{E})\;\Varid{→}\;\Conid{LazyCoercion}\;(\Varid{μ}\;\Conid{E})\;\Varid{→}\;\Conid{Set₁}{}\<[E]%
\\
\>[B]{}\hsindent{3}{}\<[3]%
\>[3]{}\Keyword{where}{}\<[E]%
\\
\>[B]{}\hsindent{3}{}\<[3]%
\>[3]{}\Varid{stepl}\;\mathbin{:}\;\{\mskip1.5mu \Varid{e₁}\;\Varid{e₁'}\;\Varid{e₂}\;\mathbin{:}\;\Varid{μ}\;\Conid{E}\mskip1.5mu\}\;{}\<[E]%
\\
\>[3]{}\hsindent{2}{}\<[5]%
\>[5]{}\Varid{→}\;\Varid{e₁}\;\Varid{⟶}\;\Varid{e₁'}\;{}\<[E]%
\\
\>[3]{}\hsindent{2}{}\<[5]%
\>[5]{}\Varid{→}\;\Varid{inj}\;\Varid{lift⁺}\;(\Varid{e₁},\Varid{e₂})\;\Varid{⟶⁺}\;\Varid{inj}\;\Varid{lift⁺}\;(\Varid{e₁'},\Varid{e₂}){}\<[E]%
\\
\>[B]{}\hsindent{3}{}\<[3]%
\>[3]{}\Varid{stepr}\;\mathbin{:}\;\{\mskip1.5mu \Varid{e₁}\;\Varid{e₂}\;\Varid{e₂'}\;\mathbin{:}\;\Varid{μ}\;\Conid{E}\mskip1.5mu\}\;{}\<[E]%
\\
\>[3]{}\hsindent{2}{}\<[5]%
\>[5]{}\Varid{→}\;\Varid{e₂}\;\Varid{⟶}\;\Varid{e₂'}\;{}\<[E]%
\\
\>[3]{}\hsindent{2}{}\<[5]%
\>[5]{}\Varid{→}\;\Varid{inj}\;\Varid{lift⁺}\;(\Varid{e₁},\Varid{e₂})\;\Varid{⟶⁺}\;\Varid{inj}\;\Varid{lift⁺}\;(\Varid{e₁},\Varid{e₂'}){}\<[E]%
\\
\>[B]{}\hsindent{3}{}\<[3]%
\>[3]{}\Varid{stepv}\;\mathbin{:}\;\{\mskip1.5mu \Varid{n}\;\Varid{m}\;\mathbin{:}\;\Conid{ℕ}\mskip1.5mu\}\;{}\<[E]%
\\
\>[3]{}\hsindent{2}{}\<[5]%
\>[5]{}\Varid{→}\;\Varid{inj}\;\Varid{lift⁺}\;(\Varid{apply}\;\Varid{liftℕ}\;\Varid{n},\Varid{apply}\;\Varid{liftℕ}\;\Varid{m})\;{}\<[E]%
\\
\>[5]{}\hsindent{2}{}\<[7]%
\>[7]{}\Varid{⟶⁺}\;\Varid{inj}\;\Varid{liftℕ}\;(\Varid{n}\;\Varid{+ℕ}\;\Varid{m}){}\<[E]%
\ColumnHook
\end{hscode}\resethooks
It appears we've littered an otherwise simple definition with $inj$ but we've
replaced our arbitrary arrows with objects having constructors we can match on.
Using the above techniques we can modularize the welltyping relation over
sums for free
\begin{hscode}\SaveRestoreHook
\column{B}{@{}>{\hspre}l<{\hspost}@{}}%
\column{3}{@{}>{\hspre}l<{\hspost}@{}}%
\column{5}{@{}>{\hspre}l<{\hspost}@{}}%
\column{E}{@{}>{\hspre}l<{\hspost}@{}}%
\>[B]{}\Keyword{data}\;\Conid{WtSum}\;\{\mskip1.5mu \Conid{E}\;\mathbin{:}\;\Conid{Functor}\mskip1.5mu\}{}\<[E]%
\\
\>[B]{}\hsindent{3}{}\<[3]%
\>[3]{}\{\mskip1.5mu \Conid{Wt}\;\mathbin{:}\;\Varid{μ}\;\Conid{E}\;\Varid{→}\;\Conid{Type}\;\Varid{→}\;\Conid{Set₁}\mskip1.5mu\}{}\<[E]%
\\
\>[B]{}\hsindent{3}{}\<[3]%
\>[3]{}\{\mskip1.5mu \Varid{lift⁺}\;\mathbin{:}\;[\mskip1.5mu \Conid{Sum}\mskip1.5mu]\;(\Varid{μ}\;\Conid{E})\;\Varid{↣}\;\Varid{μ}\;\Conid{E}\mskip1.5mu\}{}\<[E]%
\\
\>[B]{}\hsindent{3}{}\<[3]%
\>[3]{}\mathbin{:}\;\Conid{LazyCoercion}\;(\Varid{μ}\;\Conid{E})\;\Varid{→}\;\Conid{Type}\;\Varid{→}\;\Conid{Set₁}\;\Keyword{where}{}\<[E]%
\\
\>[B]{}\hsindent{3}{}\<[3]%
\>[3]{}\Varid{ok-sum}\;\mathbin{:}\;\{\mskip1.5mu \Varid{e₁}\;\Varid{e₂}\;\mathbin{:}\;\Varid{μ}\;\Conid{E}\mskip1.5mu\}\;{}\<[E]%
\\
\>[3]{}\hsindent{2}{}\<[5]%
\>[5]{}\Varid{→}\;\Conid{Wt}\;\Varid{e₁}\;\Conid{TNat}\;\Varid{→}\;\Conid{Wt}\;\Varid{e₂}\;\Conid{TNat}\;{}\<[E]%
\\
\>[3]{}\hsindent{2}{}\<[5]%
\>[5]{}\Varid{→}\;\Conid{WtSum}\;(\Varid{inj}\;\Varid{lift⁺}\;(\Varid{e₁},\Varid{e₂}))\;\Conid{TNat}{}\<[E]%
\ColumnHook
\end{hscode}\resethooks
The above definitions nearly wrote themselves. The simplicity comes from the
fact we are just abstracting as many terms as possible, keeping in mind
we can fill them in naturally later because the abstraction is so general
there are few options available.

\subsection{Arrays}
We proceed by defining the step and welltypedness relations on arrays
that can be combined with the relations on sums.
The definitions for evaluation and welltypedness should look
similar to those for sums $-\longrightarrow^{+}-$. 
\begin{hscode}\SaveRestoreHook
\column{B}{@{}>{\hspre}l<{\hspost}@{}}%
\column{3}{@{}>{\hspre}l<{\hspost}@{}}%
\column{5}{@{}>{\hspre}l<{\hspost}@{}}%
\column{9}{@{}>{\hspre}l<{\hspost}@{}}%
\column{E}{@{}>{\hspre}l<{\hspost}@{}}%
\>[B]{}\Keyword{data}\;\Varid{\char95 ⟶}\;[\mskip1.5mu \mskip1.5mu]\;\anonymous \;\{\mskip1.5mu \Conid{E}\;\mathbin{:}\;\Conid{Functor}\mskip1.5mu\}{}\<[E]%
\\
\>[B]{}\hsindent{3}{}\<[3]%
\>[3]{}\{\mskip1.5mu \Varid{\char95 ⟶\char95 }\;\mathbin{:}\;\Varid{μ}\;\Conid{E}\;\Varid{→}\;\Varid{μ}\;\Conid{E}\;\Varid{→}\;\Conid{Set₁}\mskip1.5mu\}{}\<[E]%
\\
\>[B]{}\hsindent{3}{}\<[3]%
\>[3]{}\{\mskip1.5mu \Varid{liftA}\;\mathbin{:}\;[\mskip1.5mu \Conid{Array}\mskip1.5mu]\;(\Varid{μ}\;\Conid{E})\;\Varid{↣}\;\Varid{μ}\;\Conid{E}\mskip1.5mu\}{}\<[E]%
\\
\>[B]{}\hsindent{3}{}\<[3]%
\>[3]{}\{\mskip1.5mu \Varid{liftℕ}\;\mathbin{:}\;\Conid{ℕ}\;\Varid{↣}\;\Varid{μ}\;\Conid{E}\mskip1.5mu\}{}\<[E]%
\\
\>[B]{}\hsindent{3}{}\<[3]%
\>[3]{}\{\mskip1.5mu \Varid{liftO}\;\mathbin{:}\;[\mskip1.5mu \Conid{Option}\mskip1.5mu]\;(\Varid{μ}\;\Conid{E})\;\Varid{↣}\;\Varid{μ}\;\Conid{E}\mskip1.5mu\}{}\<[E]%
\\
\>[B]{}\hsindent{3}{}\<[3]%
\>[3]{}\mathbin{:}\;\Conid{LazyCoercion}\;(\Varid{μ}\;\Conid{E})\;\Varid{→}\;\Conid{LazyCoercion}\;(\Varid{μ}\;\Conid{E})\;\Varid{→}\;\Conid{Set₁}{}\<[E]%
\\
\>[B]{}\hsindent{3}{}\<[3]%
\>[3]{}\Keyword{where}{}\<[E]%
\\
\>[B]{}\hsindent{3}{}\<[3]%
\>[3]{}\Varid{stepi}\;\mathbin{:}\;\{\mskip1.5mu \Varid{e}\;\Varid{e'}\;\Varid{a}\;\mathbin{:}\;\Varid{μ}\;\Conid{E}\mskip1.5mu\}\;\Varid{→}\;\Varid{e}\;\Varid{⟶}\;\Varid{e'}\;{}\<[E]%
\\
\>[3]{}\hsindent{2}{}\<[5]%
\>[5]{}\Varid{→}\;\Varid{inj}\;\Varid{liftA}\;(\Varid{a}\;\mathbin{!}\;\Varid{e})\;\Varid{⟶}\;[\mskip1.5mu \mskip1.5mu]\;\Varid{inj}\;\Varid{liftA}\;(\Varid{a}\;\mathbin{!}\;\Varid{e'}){}\<[E]%
\\
\>[B]{}\hsindent{3}{}\<[3]%
\>[3]{}\Varid{lookup}\;\mathbin{:}\;\{\mskip1.5mu \Varid{a}\;\mathbin{:}\;[\mskip1.5mu \Conid{Array}\mskip1.5mu]\;(\Varid{μ}\;\Conid{E})\mskip1.5mu\}\;\{\mskip1.5mu \Varid{n}\;\mathbin{:}\;\Conid{ℕ}\mskip1.5mu\}\;{}\<[E]%
\\
\>[3]{}\hsindent{2}{}\<[5]%
\>[5]{}\Varid{→}\;\Varid{inj}\;\Varid{liftA}\;(\Varid{apply}\;\Varid{liftA}\;\Varid{a}\;\mathbin{!}\;\Varid{apply}\;\Varid{liftℕ}\;\Varid{n})\;{}\<[E]%
\\
\>[5]{}\hsindent{4}{}\<[9]%
\>[9]{}\Varid{⟶}\;[\mskip1.5mu \mskip1.5mu]\;\Varid{inj}\;\Varid{liftO}\;\Conid{L⟦}\;\Varid{a},\Varid{n}\;\Varid{⟧}{}\<[E]%
\ColumnHook
\end{hscode}\resethooks
To define the typing relation we again follow the format of $\mathit{WtSum}$ 
above and we are done.
\begin{hscode}\SaveRestoreHook
\column{B}{@{}>{\hspre}l<{\hspost}@{}}%
\column{3}{@{}>{\hspre}l<{\hspost}@{}}%
\column{5}{@{}>{\hspre}l<{\hspost}@{}}%
\column{E}{@{}>{\hspre}l<{\hspost}@{}}%
\>[B]{}\Keyword{data}\;\Conid{WtArray}\;\{\mskip1.5mu \Conid{E}\;\mathbin{:}\;\Conid{Functor}\mskip1.5mu\}{}\<[E]%
\\
\>[B]{}\hsindent{3}{}\<[3]%
\>[3]{}\{\mskip1.5mu \Conid{Wt}\;\mathbin{:}\;\Varid{μ}\;\Conid{E}\;\Varid{→}\;\Conid{Type}\;\Varid{→}\;\Conid{Set₁}\mskip1.5mu\}{}\<[E]%
\\
\>[B]{}\hsindent{3}{}\<[3]%
\>[3]{}\{\mskip1.5mu \Varid{liftA}\;\mathbin{:}\;[\mskip1.5mu \Conid{Array}\mskip1.5mu]\;(\Varid{μ}\;\Conid{E})\;\Varid{↣}\;(\Varid{μ}\;\Conid{E})\mskip1.5mu\}{}\<[E]%
\\
\>[B]{}\hsindent{3}{}\<[3]%
\>[3]{}\{\mskip1.5mu \Varid{liftℕ}\;\mathbin{:}\;\Conid{ℕ}\;\Varid{↣}\;\Varid{μ}\;\Conid{E}\mskip1.5mu\}{}\<[E]%
\\
\>[B]{}\hsindent{3}{}\<[3]%
\>[3]{}\mathbin{:}\;\Conid{LazyCoercion}\;(\Varid{μ}\;\Conid{E})\;\Varid{→}\;\Conid{Type}\;\Varid{→}\;\Conid{Set₁}\;\Keyword{where}{}\<[E]%
\\
\>[B]{}\hsindent{3}{}\<[3]%
\>[3]{}\Varid{ok-nil}\;\mathbin{:}\;\Conid{WtArray}\;(\Varid{inj}\;\Varid{liftA}\;\Varid{nil})\;\Conid{TArray}{}\<[E]%
\\
\>[B]{}\hsindent{3}{}\<[3]%
\>[3]{}\Varid{ok-ins}\;\mathbin{:}\;\{\mskip1.5mu \Varid{a}\;\Varid{e}\;\Varid{n}\;\mathbin{:}\;\Varid{μ}\;\Conid{E}\mskip1.5mu\}\;{}\<[E]%
\\
\>[3]{}\hsindent{2}{}\<[5]%
\>[5]{}\Varid{→}\;\Conid{Wt}\;\Varid{a}\;\Conid{TArray}\;\Varid{→}\;\Conid{Wt}\;\Varid{e}\;\Conid{TNat}\;\Varid{→}\;\Conid{Wt}\;\Varid{n}\;\Conid{TNat}\;{}\<[E]%
\\
\>[3]{}\hsindent{2}{}\<[5]%
\>[5]{}\Varid{→}\;\Conid{WtArray}\;(\Varid{inj}\;\Varid{liftA}\;(\Varid{a}\;[\mskip1.5mu \Varid{n}\mskip1.5mu]\;\Conid{:=}\;\Varid{e}))\;\Conid{TArray}{}\<[E]%
\\
\>[B]{}\hsindent{3}{}\<[3]%
\>[3]{}\Varid{ok-lookup}\;\mathbin{:}\;\{\mskip1.5mu \Varid{e}\;\Varid{a}\;\mathbin{:}\;\Varid{μ}\;\Conid{E}\mskip1.5mu\}\;{}\<[E]%
\\
\>[3]{}\hsindent{2}{}\<[5]%
\>[5]{}\Varid{→}\;\Conid{Wt}\;\Varid{a}\;\Conid{TArray}\;\Varid{→}\;\Conid{Wt}\;\Varid{e}\;\Conid{TNat}\;{}\<[E]%
\\
\>[3]{}\hsindent{2}{}\<[5]%
\>[5]{}\Varid{→}\;\Conid{WtArray}\;(\Varid{inj}\;\Varid{liftA}\;(\Varid{a}\;\mathbin{!}\;\Varid{e}))\;\Conid{TOption}{}\<[E]%
\ColumnHook
\end{hscode}\resethooks

\section{Proving Type Preservation}
The type preservation lemma states that if a term is welltyped and can step,
then the type of the term is preserved after evaluation
\begin{equation}
  \tag{type-preservation}
  e \longrightarrow e' \wedge e : T \Rightarrow e' : T
  \label{eqn:type-preservation}
\end{equation}
Prior to considering how type preservation might look for each of the previously
defined components we should review what type preservation looks like for
the $\mathit{MonolithicExpr}$ language.  The proof is standard, proceeding by structural
induction on the shape of the welltyping tree.
\begin{hscode}\SaveRestoreHook
\column{B}{@{}>{\hspre}l<{\hspost}@{}}%
\column{3}{@{}>{\hspre}l<{\hspost}@{}}%
\column{E}{@{}>{\hspre}l<{\hspost}@{}}%
\>[B]{}\Varid{preservation-MonolithicExpr}\;\mathbin{:}\;\Varid{∀}\;\{\mskip1.5mu \Varid{e}\;\Varid{e'}\mskip1.5mu\}\;\{\mskip1.5mu \Varid{τ}\mskip1.5mu\}\;{}\<[E]%
\\
\>[B]{}\hsindent{3}{}\<[3]%
\>[3]{}\Varid{→}\;\Varid{e}\;\Varid{⟶C}\;\Varid{e'}\;{}\<[E]%
\\
\>[B]{}\hsindent{3}{}\<[3]%
\>[3]{}\Varid{→}\;\Conid{WtMonolithicExpr}\;\Varid{e}\;\Varid{τ}\;{}\<[E]%
\\
\>[B]{}\hsindent{3}{}\<[3]%
\>[3]{}\Varid{→}\;\Conid{WtMonolithicExpr}\;\Varid{e'}\;\Varid{τ}{}\<[E]%
\\
\>[B]{}\Varid{preservation-MonolithicExpr}\;(\Varid{stepl}\;\Varid{ste₁})\;(\Varid{ok-sum}\;\Varid{wte₁}\;\Varid{wte₂})\;{}\<[E]%
\\
\>[B]{}\hsindent{3}{}\<[3]%
\>[3]{}\mathrel{=}\;\Varid{ok-sum}\;(\Varid{preservation-MonolithicExpr}\;\Varid{ste₁}\;\Varid{wte₁})\;\Varid{wte₂}{}\<[E]%
\\
\>[B]{}\Varid{preservation-MonolithicExpr}\;(\Varid{stepr}\;\Varid{ste₂})\;(\Varid{ok-sum}\;\Varid{wte₁}\;\Varid{wte₂})\;{}\<[E]%
\\
\>[B]{}\hsindent{3}{}\<[3]%
\>[3]{}\mathrel{=}\;\Varid{ok-sum}\;\Varid{wte₁}\;(\Varid{preservation-MonolithicExpr}\;\Varid{ste₂}\;\Varid{wte₂}){}\<[E]%
\\
\>[B]{}\Varid{preservation-MonolithicExpr}\;{}\<[E]%
\\
\>[B]{}\hsindent{3}{}\<[3]%
\>[3]{}(\Varid{stepv}\;\{\mskip1.5mu \Varid{n}\mskip1.5mu\}\;\{\mskip1.5mu \Varid{m}\mskip1.5mu\})\;(\Varid{ok-sum}\;\Varid{wtn}\;\Varid{wtm})\;{}\<[E]%
\\
\>[B]{}\hsindent{3}{}\<[3]%
\>[3]{}\mathrel{=}\;\Varid{ok-nat}\;(\Varid{n}\;\Varid{+ℕ}\;\Varid{m}){}\<[E]%
\\
\>[B]{}\Varid{preservation-MonolithicExpr}\;(\Varid{stepi}\;\Varid{ste})\;(\Varid{ok-lookup}\;\Varid{wta}\;\Varid{wte})\;{}\<[E]%
\\
\>[B]{}\hsindent{3}{}\<[3]%
\>[3]{}\mathrel{=}\;\Varid{ok-lookup}\;\Varid{wta}\;(\Varid{preservation-MonolithicExpr}\;\Varid{ste}\;\Varid{wte}){}\<[E]%
\\
\>[B]{}\Varid{preservation-MonolithicExpr}\;{}\<[E]%
\\
\>[B]{}\hsindent{3}{}\<[3]%
\>[3]{}(\Varid{lookup}\;\{\mskip1.5mu \Varid{a}\mskip1.5mu\}\;\{\mskip1.5mu \Varid{n}\mskip1.5mu\})\;(\Varid{ok-lookup}\;\Varid{wta}\;\Varid{wtn})\;{}\<[E]%
\\
\>[B]{}\hsindent{3}{}\<[3]%
\>[3]{}\mathrel{=}\;\Varid{proj₂}\;\Conid{LC⟦}\;\Varid{a},\Varid{n}\;\Varid{⟧}{}\<[E]%
\ColumnHook
\end{hscode}\resethooks
There are three items worth noting here: the first is the use of the function
$LC\llbracket-,-\rrbracket : \mathit{MonolithicExpr} \rightarrow \mathbb{N}
  \rightarrow \exists e . \mathit{WtMonolithicExpr} \; e \; \mathit{TOption}$ which we have assumed
produces a pair with first component an expression and second component a
proof that the expression is a welltyped option; the second is that recursion
acts as our induction hypothesis; and finally that Agda is smart enough to
notice there is only a single possible welltyping constructor for each step
constructor---in Agda all functions are total.

We should expect the modular type preservation lemmas to look similar because
there is little global knowledge involved.  The induction hypothesis
and values aside, each case is ``contained within its own world'' in the
sense that each evaluation rule relies only on the fact that subterms are
welltyped but ignoring the \emph{reason} they are welltyped.  To show type
preservation for sums we might start with\begin{hscode}\SaveRestoreHook
\column{B}{@{}>{\hspre}l<{\hspost}@{}}%
\column{3}{@{}>{\hspre}l<{\hspost}@{}}%
\column{5}{@{}>{\hspre}l<{\hspost}@{}}%
\column{E}{@{}>{\hspre}l<{\hspost}@{}}%
\>[3]{}\Varid{preservation-Sum₁}\;\mathbin{:}\;\{\mskip1.5mu \Varid{τ}\;\mathbin{:}\;\Conid{Type}\mskip1.5mu\}\;\{\mskip1.5mu \Conid{E}\;\mathbin{:}\;\Conid{Functor}\mskip1.5mu\}\;{}\<[E]%
\\
\>[3]{}\hsindent{2}{}\<[5]%
\>[5]{}\{\mskip1.5mu \Varid{e}\;\Varid{e'}\;\mathbin{:}\;\Conid{LazyCoercion}\;(\Varid{μ}\;\Conid{E})\mskip1.5mu\}\;{}\<[E]%
\\
\>[3]{}\hsindent{2}{}\<[5]%
\>[5]{}\Varid{→}\;\Varid{e}\;\Varid{⟶⁺}\;\Varid{e'}\;{}\<[E]%
\\
\>[3]{}\hsindent{2}{}\<[5]%
\>[5]{}\Varid{→}\;\Conid{WtSum}\;\Varid{e}\;\Varid{τ}\;{}\<[E]%
\\
\>[3]{}\hsindent{2}{}\<[5]%
\>[5]{}\Varid{→}\;\Conid{WtSum}\;\Varid{e'}\;\Varid{τ}{}\<[E]%
\\
\>[3]{}\Varid{preservation-Sum₁}\;{}\<[E]%
\\
\>[3]{}\hsindent{2}{}\<[5]%
\>[5]{}(\Varid{stepl}\;\{\mskip1.5mu \Varid{e₁}\mskip1.5mu\}\;\{\mskip1.5mu \Varid{e₁'}\mskip1.5mu\}\;\{\mskip1.5mu \Varid{e₂}\mskip1.5mu\}\;\Varid{ste₁})\;(\Varid{ok-sum}\;\Varid{wte₁}\;\Varid{wte₂})\;\mathrel{=}\;\Varid{*}{}\<[E]%
\ColumnHook
\end{hscode}\resethooks
however recall that $-\longrightarrow^{+}-$ requires the top-level step
relation and proof that $E$ contains both sums and naturals. There is a
second mistake in writing preservation this way---we would like to show
that $e'$ is welltyped in the expression language, not just necessarily
in the modular sum language, this reflects our desire to expose as
little about each component as possible.  A second formulation might then begin
as follows but we again fail.  \begin{hscode}\SaveRestoreHook
\column{B}{@{}>{\hspre}l<{\hspost}@{}}%
\column{3}{@{}>{\hspre}l<{\hspost}@{}}%
\column{5}{@{}>{\hspre}l<{\hspost}@{}}%
\column{E}{@{}>{\hspre}l<{\hspost}@{}}%
\>[3]{}\Varid{preservation-Sum₂}\;\mathbin{:}\;\{\mskip1.5mu \Varid{τ}\;\mathbin{:}\;\Conid{Type}\mskip1.5mu\}\;{}\<[E]%
\\
\>[3]{}\hsindent{2}{}\<[5]%
\>[5]{}\{\mskip1.5mu \Conid{E}\;\mathbin{:}\;\Conid{Functor}\mskip1.5mu\}\;{}\<[E]%
\\
\>[3]{}\hsindent{2}{}\<[5]%
\>[5]{}\{\mskip1.5mu \Varid{\char95 ⟶\char95 }\;\mathbin{:}\;\Varid{μ}\;\Conid{E}\;\Varid{→}\;\Varid{μ}\;\Conid{E}\;\Varid{→}\;\Conid{Set₁}\mskip1.5mu\}\;{}\<[E]%
\\
\>[3]{}\hsindent{2}{}\<[5]%
\>[5]{}\{\mskip1.5mu \Varid{lift⁺}\;\mathbin{:}\;[\mskip1.5mu \Conid{Sum}\mskip1.5mu]\;(\Varid{μ}\;\Conid{E})\;\Varid{↣}\;\Varid{μ}\;\Conid{E}\mskip1.5mu\}\;{}\<[E]%
\\
\>[3]{}\hsindent{2}{}\<[5]%
\>[5]{}\{\mskip1.5mu \Varid{liftℕ}\;\mathbin{:}\;\Conid{ℕ}\;\Varid{↣}\;\Varid{μ}\;\Conid{E}\mskip1.5mu\}\;{}\<[E]%
\\
\>[3]{}\hsindent{2}{}\<[5]%
\>[5]{}\{\mskip1.5mu \Conid{Wt}\;\mathbin{:}\;\Varid{μ}\;\Conid{E}\;\Varid{→}\;\Conid{Type}\;\Varid{→}\;\Conid{Set₁}\mskip1.5mu\}\;{}\<[E]%
\\
\>[3]{}\hsindent{2}{}\<[5]%
\>[5]{}\{\mskip1.5mu \Varid{e}\;\Varid{e'}\;\mathbin{:}\;\Conid{LazyCoercion}\;(\Varid{μ}\;\Conid{E})\mskip1.5mu\}\;{}\<[E]%
\\
\>[3]{}\hsindent{2}{}\<[5]%
\>[5]{}\Varid{→}\;\Varid{\char95 ⟶⁺\char95 }\;\{\mskip1.5mu \Conid{E}\mskip1.5mu\}\;\{\mskip1.5mu \Varid{\char95 ⟶\char95 }\mskip1.5mu\}\;\{\mskip1.5mu \Varid{lift⁺}\mskip1.5mu\}\;\{\mskip1.5mu \Varid{liftℕ}\mskip1.5mu\}\;\Varid{e}\;\Varid{e'}\;{}\<[E]%
\\
\>[3]{}\hsindent{2}{}\<[5]%
\>[5]{}\Varid{→}\;\Conid{WtSum}\;\{\mskip1.5mu \Conid{E}\mskip1.5mu\}\;\{\mskip1.5mu \Conid{Wt}\mskip1.5mu\}\;\{\mskip1.5mu \Varid{lift⁺}\mskip1.5mu\}\;\Varid{e}\;\Varid{τ}\;{}\<[E]%
\\
\>[3]{}\hsindent{2}{}\<[5]%
\>[5]{}\Varid{→}\;\Conid{Wt}\;(\Varid{coerce}\;\Varid{e'})\;\Varid{τ}{}\<[E]%
\\
\>[3]{}\Varid{preservation-Sum₂}\;(\Varid{stepl}\;\Varid{ste₁})\;(\Varid{ok-sum}\;\Varid{wte₁}\;\Varid{wte₂})\;{}\<[E]%
\\
\>[3]{}\hsindent{2}{}\<[5]%
\>[5]{}\mathrel{=}\;\Varid{*}\;(\Varid{ok-sum}\;\Varid{*}\;\Varid{wte₂}){}\<[E]%
\\
\>[3]{}\Varid{preservation-Sum₂}\;(\Varid{stepr}\;\Varid{ste₁})\;(\Varid{ok-sum}\;\Varid{wte₁}\;\Varid{wte₂})\;{}\<[E]%
\\
\>[3]{}\hsindent{2}{}\<[5]%
\>[5]{}\mathrel{=}\;\Varid{*}\;(\Varid{ok-sum}\;\Varid{wte₁}\;\Varid{*}){}\<[E]%
\\
\>[3]{}\Varid{preservation-Sum₂}\;\Varid{stepv}\;(\Varid{ok-sum}\;\Varid{wte₁}\;\Varid{wte₂})\;{}\<[E]%
\\
\>[3]{}\hsindent{2}{}\<[5]%
\>[5]{}\mathrel{=}\;\Varid{*}\;(\Varid{n}\;\Varid{+ℕ}\;\Varid{m}){}\<[E]%
\ColumnHook
\end{hscode}\resethooks
It seems we're only missing two pieces:
we need to be able to lift welltyped sums and naturals into $\mathit{Wt}$;
and we need some way of expressing the induction hypothesis which states
that because $e_1$ is welltyped and stepped, $e_1'$ is welltyped too.
The induction hypothesis is slightly stranger than was the case in our
$\mathit{MonolithicExpr}$'s because we know $e_1$ and $e_1'$ are welltyped
despite the fact that they are any expressions, not necessarily just sums.
This motivates our solution which takes the induction hypothesis as an
explicit assumption.
\begin{hscode}\SaveRestoreHook
\column{B}{@{}>{\hspre}l<{\hspost}@{}}%
\column{3}{@{}>{\hspre}l<{\hspost}@{}}%
\column{5}{@{}>{\hspre}l<{\hspost}@{}}%
\column{E}{@{}>{\hspre}l<{\hspost}@{}}%
\>[B]{}\Varid{preservation-Sum}\;\mathbin{:}\;\{\mskip1.5mu \Varid{τ}\;\mathbin{:}\;\Conid{Type}\mskip1.5mu\}\;{}\<[E]%
\\
\>[B]{}\hsindent{3}{}\<[3]%
\>[3]{}\{\mskip1.5mu \Conid{E}\;\mathbin{:}\;\Conid{Functor}\mskip1.5mu\}\;{}\<[E]%
\\
\>[B]{}\hsindent{3}{}\<[3]%
\>[3]{}\{\mskip1.5mu \Varid{\char95 ⟶\char95 }\;\mathbin{:}\;\Varid{μ}\;\Conid{E}\;\Varid{→}\;\Varid{μ}\;\Conid{E}\;\Varid{→}\;\Conid{Set₁}\mskip1.5mu\}\;{}\<[E]%
\\
\>[B]{}\hsindent{3}{}\<[3]%
\>[3]{}\{\mskip1.5mu \Varid{lift⁺}\;\mathbin{:}\;[\mskip1.5mu \Conid{Sum}\mskip1.5mu]\;(\Varid{μ}\;\Conid{E})\;\Varid{↣}\;\Varid{μ}\;\Conid{E}\mskip1.5mu\}\;{}\<[E]%
\\
\>[B]{}\hsindent{3}{}\<[3]%
\>[3]{}\{\mskip1.5mu \Varid{liftℕ}\;\mathbin{:}\;\Conid{ℕ}\;\Varid{↣}\;\Varid{μ}\;\Conid{E}\mskip1.5mu\}\;{}\<[E]%
\\
\>[B]{}\hsindent{3}{}\<[3]%
\>[3]{}\{\mskip1.5mu \Conid{Wt}\;\mathbin{:}\;\Varid{μ}\;\Conid{E}\;\Varid{→}\;\Conid{Type}\;\Varid{→}\;\Conid{Set₁}\mskip1.5mu\}\;{}\<[E]%
\\
\>[B]{}\hsindent{3}{}\<[3]%
\>[3]{}\{\mskip1.5mu \Varid{a}\;\Varid{b}\;\mathbin{:}\;\Conid{LazyCoercion}\;(\Varid{μ}\;\Conid{E})\mskip1.5mu\}\;{}\<[E]%
\\
\>[B]{}\hsindent{3}{}\<[3]%
\>[3]{}\Varid{→}\;((\Varid{n}\;\mathbin{:}\;\Conid{ℕ})\;\Varid{→}\;\Conid{Wt}\;(\Varid{apply}\;\Varid{liftℕ}\;\Varid{n})\;\Conid{TNat})\;{}\<[E]%
\\
\>[B]{}\hsindent{3}{}\<[3]%
\>[3]{}\Varid{→}\;(\Varid{∀}\;\{\mskip1.5mu \Varid{δ}\mskip1.5mu\}\;\{\mskip1.5mu \Varid{e}\mskip1.5mu\}{}\<[E]%
\\
\>[3]{}\hsindent{2}{}\<[5]%
\>[5]{}\Varid{→}\;\Conid{WtSum}\;\{\mskip1.5mu \Conid{E}\mskip1.5mu\}\;\{\mskip1.5mu \Conid{Wt}\mskip1.5mu\}\;\{\mskip1.5mu \Varid{lift⁺}\mskip1.5mu\}\;(\Varid{inj}\;\Varid{lift⁺}\;\Varid{e})\;\Varid{δ}{}\<[E]%
\\
\>[3]{}\hsindent{2}{}\<[5]%
\>[5]{}\Varid{→}\;\Conid{Wt}\;(\Varid{apply}\;\Varid{lift⁺}\;\Varid{e})\;\Varid{δ})\;{}\<[E]%
\\
\>[B]{}\hsindent{3}{}\<[3]%
\>[3]{}\Varid{→}\;(\Varid{∀}\;\{\mskip1.5mu \Varid{δ}\mskip1.5mu\}\;\{\mskip1.5mu \Varid{e}\;\Varid{e'}\mskip1.5mu\}\;\Varid{→}\;\Varid{e}\;\Varid{⟶}\;\Varid{e'}\;\Varid{→}\;\Conid{Wt}\;\Varid{e}\;\Varid{δ}\;\Varid{→}\;\Conid{Wt}\;\Varid{e'}\;\Varid{δ})\;{}\<[E]%
\\
\>[B]{}\hsindent{3}{}\<[3]%
\>[3]{}\Varid{→}\;\Varid{\char95 ⟶⁺\char95 }\;\{\mskip1.5mu \Conid{E}\mskip1.5mu\}\;\{\mskip1.5mu \Varid{\char95 ⟶\char95 }\mskip1.5mu\}\;\{\mskip1.5mu \Varid{lift⁺}\mskip1.5mu\}\;\{\mskip1.5mu \Varid{liftℕ}\mskip1.5mu\}\;\Varid{a}\;\Varid{b}\;{}\<[E]%
\\
\>[B]{}\hsindent{3}{}\<[3]%
\>[3]{}\Varid{→}\;\Conid{WtSum}\;\{\mskip1.5mu \Conid{E}\mskip1.5mu\}\;\{\mskip1.5mu \Conid{Wt}\mskip1.5mu\}\;\{\mskip1.5mu \Varid{lift⁺}\mskip1.5mu\}\;\Varid{a}\;\Varid{τ}\;{}\<[E]%
\\
\>[B]{}\hsindent{3}{}\<[3]%
\>[3]{}\Varid{→}\;\Conid{Wt}\;(\Varid{coerce}\;\Varid{b})\;\Varid{τ}{}\<[E]%
\\
\>[B]{}\Varid{preservation-Sum}\;\Varid{wtnat}\;\Varid{wt}\;\Conid{IH}\;{}\<[E]%
\\
\>[B]{}\hsindent{3}{}\<[3]%
\>[3]{}(\Varid{stepl}\;\Varid{ste₁})\;(\Varid{ok-sum}\;\Varid{wte₁}\;\Varid{wte₂})\;{}\<[E]%
\\
\>[B]{}\hsindent{3}{}\<[3]%
\>[3]{}\mathrel{=}\;\Varid{wt}\;(\Varid{ok-sum}\;(\Conid{IH}\;\Varid{ste₁}\;\Varid{wte₁})\;\Varid{wte₂}){}\<[E]%
\\
\>[B]{}\Varid{preservation-Sum}\;\Varid{wtnat}\;\Varid{wt}\;\Conid{IH}\;{}\<[E]%
\\
\>[B]{}\hsindent{3}{}\<[3]%
\>[3]{}(\Varid{stepr}\;\Varid{ste₂})\;(\Varid{ok-sum}\;\Varid{wte₁}\;\Varid{wte₂})\;{}\<[E]%
\\
\>[B]{}\hsindent{3}{}\<[3]%
\>[3]{}\mathrel{=}\;\Varid{wt}\;(\Varid{ok-sum}\;\Varid{wte₁}\;(\Conid{IH}\;\Varid{ste₂}\;\Varid{wte₂})){}\<[E]%
\\
\>[B]{}\Varid{preservation-Sum}\;\Varid{wtnat}\;\Varid{wt}\;\Conid{IH}\;{}\<[E]%
\\
\>[B]{}\hsindent{3}{}\<[3]%
\>[3]{}(\Varid{stepv}\;\{\mskip1.5mu \Varid{n}\mskip1.5mu\}\;\{\mskip1.5mu \Varid{m}\mskip1.5mu\})\;(\Varid{ok-sum}\;\Varid{wte₁}\;\Varid{wte₂})\;{}\<[E]%
\\
\>[B]{}\hsindent{3}{}\<[3]%
\>[3]{}\mathrel{=}\;\Varid{wtnat}\;(\Varid{n}\;\Varid{+ℕ}\;\Varid{m}){}\<[E]%
\ColumnHook
\end{hscode}\resethooks
We are pleased with how similar this is to the original, monolithic formulation.
Notice again that the solution was to factor out assumptions about the outside
world similar to the previous abstractions.  Proving type preservation for
arrays is similarly natural:
\begin{hscode}\SaveRestoreHook
\column{B}{@{}>{\hspre}l<{\hspost}@{}}%
\column{3}{@{}>{\hspre}l<{\hspost}@{}}%
\column{5}{@{}>{\hspre}l<{\hspost}@{}}%
\column{E}{@{}>{\hspre}l<{\hspost}@{}}%
\>[B]{}\Varid{preservation-Array}\;\mathbin{:}\;\{\mskip1.5mu \Varid{τ}\;\mathbin{:}\;\Conid{Type}\mskip1.5mu\}\;{}\<[E]%
\\
\>[B]{}\hsindent{3}{}\<[3]%
\>[3]{}\{\mskip1.5mu \Conid{E}\;\mathbin{:}\;\Conid{Functor}\mskip1.5mu\}\;{}\<[E]%
\\
\>[B]{}\hsindent{3}{}\<[3]%
\>[3]{}\{\mskip1.5mu \Varid{\char95 ⟶\char95 }\;\mathbin{:}\;\Varid{μ}\;\Conid{E}\;\Varid{→}\;\Varid{μ}\;\Conid{E}\;\Varid{→}\;\Conid{Set₁}\mskip1.5mu\}\;{}\<[E]%
\\
\>[B]{}\hsindent{3}{}\<[3]%
\>[3]{}\{\mskip1.5mu \Varid{liftA}\;\mathbin{:}\;[\mskip1.5mu \Conid{Array}\mskip1.5mu]\;(\Varid{μ}\;\Conid{E})\;\Varid{↣}\;(\Varid{μ}\;\Conid{E})\mskip1.5mu\}\;{}\<[E]%
\\
\>[B]{}\hsindent{3}{}\<[3]%
\>[3]{}\{\mskip1.5mu \Varid{liftℕ}\;\mathbin{:}\;\Conid{ℕ}\;\Varid{↣}\;\Varid{μ}\;\Conid{E}\mskip1.5mu\}\;{}\<[E]%
\\
\>[B]{}\hsindent{3}{}\<[3]%
\>[3]{}\{\mskip1.5mu \Varid{liftO}\;\mathbin{:}\;[\mskip1.5mu \Conid{Option}\mskip1.5mu]\;(\Varid{μ}\;\Conid{E})\;\Varid{↣}\;(\Varid{μ}\;\Conid{E})\mskip1.5mu\}\;{}\<[E]%
\\
\>[B]{}\hsindent{3}{}\<[3]%
\>[3]{}\{\mskip1.5mu \Conid{Wt}\;\mathbin{:}\;\Varid{μ}\;\Conid{E}\;\Varid{→}\;\Conid{Type}\;\Varid{→}\;\Conid{Set₁}\mskip1.5mu\}\;{}\<[E]%
\\
\>[B]{}\hsindent{3}{}\<[3]%
\>[3]{}\{\mskip1.5mu \Varid{a}\;\Varid{b}\;\mathbin{:}\;\Conid{LazyCoercion}\;(\Varid{μ}\;\Conid{E})\mskip1.5mu\}\;{}\<[E]%
\\
\>[B]{}\hsindent{3}{}\<[3]%
\>[3]{}\Varid{→}\;((\Varid{m}\;\mathbin{:}\;[\mskip1.5mu \Conid{Option}\mskip1.5mu]\;(\Varid{μ}\;\Conid{E}))\;\Varid{→}\;\Conid{Wt}\;(\Varid{apply}\;\Varid{liftO}\;\Varid{m})\;\Conid{TOption})\;{}\<[E]%
\\
\>[B]{}\hsindent{3}{}\<[3]%
\>[3]{}\Varid{→}\;(\Varid{∀}\;\{\mskip1.5mu \Varid{δ}\mskip1.5mu\}\;\{\mskip1.5mu \Varid{e}\mskip1.5mu\}{}\<[E]%
\\
\>[3]{}\hsindent{2}{}\<[5]%
\>[5]{}\Varid{→}\;\Conid{WtArray}\;\{\mskip1.5mu \Conid{E}\mskip1.5mu\}\;\{\mskip1.5mu \Conid{Wt}\mskip1.5mu\}\;\{\mskip1.5mu \Varid{liftA}\mskip1.5mu\}\;\{\mskip1.5mu \Varid{liftℕ}\mskip1.5mu\}\;(\Varid{inj}\;\Varid{liftA}\;\Varid{e})\;\Varid{δ}{}\<[E]%
\\
\>[3]{}\hsindent{2}{}\<[5]%
\>[5]{}\Varid{→}\;\Conid{Wt}\;(\Varid{apply}\;\Varid{liftA}\;\Varid{e})\;\Varid{δ})\;{}\<[E]%
\\
\>[B]{}\hsindent{3}{}\<[3]%
\>[3]{}\Varid{→}\;(\Varid{∀}\;\{\mskip1.5mu \Varid{δ}\mskip1.5mu\}\;\{\mskip1.5mu \Varid{e}\;\Varid{e'}\mskip1.5mu\}\;\Varid{→}\;\Varid{e}\;\Varid{⟶}\;\Varid{e'}\;\Varid{→}\;\Conid{Wt}\;\Varid{e}\;\Varid{δ}\;\Varid{→}\;\Conid{Wt}\;\Varid{e'}\;\Varid{δ})\;{}\<[E]%
\\
\>[B]{}\hsindent{3}{}\<[3]%
\>[3]{}\Varid{→}\;\Varid{\char95 ⟶}\;[\mskip1.5mu \mskip1.5mu]\;\anonymous \;\{\mskip1.5mu \Conid{E}\mskip1.5mu\}\;\{\mskip1.5mu \Varid{\char95 ⟶\char95 }\mskip1.5mu\}\;\{\mskip1.5mu \Varid{liftA}\mskip1.5mu\}\;\{\mskip1.5mu \Varid{liftℕ}\mskip1.5mu\}\;\{\mskip1.5mu \Varid{liftO}\mskip1.5mu\}\;\Varid{a}\;\Varid{b}\;{}\<[E]%
\\
\>[B]{}\hsindent{3}{}\<[3]%
\>[3]{}\Varid{→}\;\Conid{WtArray}\;\{\mskip1.5mu \Conid{E}\mskip1.5mu\}\;\{\mskip1.5mu \Conid{Wt}\mskip1.5mu\}\;\{\mskip1.5mu \Varid{liftA}\mskip1.5mu\}\;\{\mskip1.5mu \Varid{liftℕ}\mskip1.5mu\}\;\Varid{a}\;\Varid{τ}\;{}\<[E]%
\\
\>[B]{}\hsindent{3}{}\<[3]%
\>[3]{}\Varid{→}\;\Conid{Wt}\;(\Varid{coerce}\;\Varid{b})\;\Varid{τ}{}\<[E]%
\\
\>[B]{}\Varid{preservation-Array}\;\Varid{wtopt}\;\Varid{wt}\;\Conid{IH}\;{}\<[E]%
\\
\>[B]{}\hsindent{3}{}\<[3]%
\>[3]{}(\Varid{stepi}\;\Varid{ste})\;(\Varid{ok-lookup}\;\Varid{wta}\;\Varid{wte})\;{}\<[E]%
\\
\>[B]{}\hsindent{3}{}\<[3]%
\>[3]{}\mathrel{=}\;\Varid{wt}\;(\Varid{ok-lookup}\;\Varid{wta}\;(\Conid{IH}\;\Varid{ste}\;\Varid{wte})){}\<[E]%
\\
\>[B]{}\Varid{preservation-Array}\;\Varid{wtopt}\;\Varid{wt}\;\Conid{IH}\;{}\<[E]%
\\
\>[B]{}\hsindent{3}{}\<[3]%
\>[3]{}(\Varid{lookup}\;\{\mskip1.5mu \Varid{a}\mskip1.5mu\}\;\{\mskip1.5mu \Varid{n}\mskip1.5mu\})\;(\Varid{ok-lookup}\;\Varid{wta}\;\Varid{wte})\;{}\<[E]%
\\
\>[B]{}\hsindent{3}{}\<[3]%
\>[3]{}\mathrel{=}\;\Varid{wtopt}\;\Conid{L⟦}\;\Varid{a},\Varid{n}\;\Varid{⟧}{}\<[E]%
\ColumnHook
\end{hscode}\resethooks
It would seem we're nearly done and the final pieces should be entirely guided
by the selected abstractions.  The $\mathit{lift}$ functions each have a unique solution:
\begin{hscode}\SaveRestoreHook
\column{B}{@{}>{\hspre}l<{\hspost}@{}}%
\column{E}{@{}>{\hspre}l<{\hspost}@{}}%
\>[B]{}\Varid{lift⁺}\;\mathbin{:}\;[\mskip1.5mu \Conid{Sum}\mskip1.5mu]\;\Conid{Expr}\;\Varid{↣}\;\Conid{Expr}{}\<[E]%
\\
\>[B]{}\Varid{lift⁺}\;\mathrel{=}\;\Varid{inj}\;(\Varid{right}\;(\Varid{left}\;(\Varid{refl}))){}\<[E]%
\\[\blanklineskip]%
\>[B]{}\Varid{liftℕ}\;\mathbin{:}\;\Conid{ℕ}\;\Varid{↣}\;\Conid{Expr}{}\<[E]%
\\
\>[B]{}\Varid{liftℕ}\;\mathrel{=}\;\Varid{inj}\;(\Varid{left}\;(\Varid{left}\;(\Varid{left}\;\Varid{refl}))){}\<[E]%
\\[\blanklineskip]%
\>[B]{}\Varid{liftO}\;\mathbin{:}\;[\mskip1.5mu \Conid{Option}\mskip1.5mu]\;\Conid{Expr}\;\Varid{↣}\;\Conid{Expr}{}\<[E]%
\\
\>[B]{}\Varid{liftO}\;\mathrel{=}\;\Varid{inj}\;(\Varid{right}\;(\Varid{left}\;(\Varid{left}\;\Varid{refl}))){}\<[E]%
\\[\blanklineskip]%
\>[B]{}\Varid{liftA}\;\mathbin{:}\;[\mskip1.5mu \Conid{Array}\mskip1.5mu]\;\Conid{Expr}\;\Varid{↣}\;\Conid{Expr}{}\<[E]%
\\
\>[B]{}\Varid{liftA}\;\mathrel{=}\;\Varid{inj}\;(\Varid{right}\;\Varid{refl}){}\<[E]%
\ColumnHook
\end{hscode}\resethooks
But how should we define welltypedness for $\mathit{Expr}$? 
Again the notion of what it
means to be welltyped has already been defined and we simply need to
``tie the knot'' as $\mathit{RecExpr}$ did above
\begin{hscode}\SaveRestoreHook
\column{B}{@{}>{\hspre}l<{\hspost}@{}}%
\column{3}{@{}>{\hspre}l<{\hspost}@{}}%
\column{5}{@{}>{\hspre}l<{\hspost}@{}}%
\column{9}{@{}>{\hspre}l<{\hspost}@{}}%
\column{16}{@{}>{\hspre}l<{\hspost}@{}}%
\column{E}{@{}>{\hspre}l<{\hspost}@{}}%
\>[B]{}\Keyword{data}\;\Conid{WtExpr}\;\mathbin{:}\;\Conid{Expr}\;\Varid{→}\;\Conid{Type}\;\Varid{→}\;\Conid{Set₁}\;\Keyword{where}{}\<[E]%
\\
\>[B]{}\hsindent{3}{}\<[3]%
\>[3]{}\Varid{lift-wt-nat}\;{}\<[16]%
\>[16]{}\mathbin{:}\;(\Varid{n}\;\mathbin{:}\;\Conid{ℕ})\;\Varid{→}\;\Conid{WtExpr}\;(\Varid{apply}\;\Varid{liftℕ}\;\Varid{n})\;\Conid{TNat}{}\<[E]%
\\
\>[B]{}\hsindent{3}{}\<[3]%
\>[3]{}\Varid{lift-wt-option}\;\mathbin{:}\;(\Varid{m}\;\mathbin{:}\;[\mskip1.5mu \Conid{Option}\mskip1.5mu]\;\Conid{Expr})\;{}\<[E]%
\\
\>[3]{}\hsindent{2}{}\<[5]%
\>[5]{}\Varid{→}\;\Conid{WtExpr}\;(\Varid{apply}\;\Varid{liftO}\;\Varid{m})\;\Conid{TOption}{}\<[E]%
\\
\>[B]{}\hsindent{3}{}\<[3]%
\>[3]{}\Varid{lift-wt-sum}\;\mathbin{:}\;\{\mskip1.5mu \Varid{τ}\;\mathbin{:}\;\Conid{Type}\mskip1.5mu\}\;\{\mskip1.5mu \Varid{e}\;\mathbin{:}\;[\mskip1.5mu \Conid{Sum}\mskip1.5mu]\;\Conid{Expr}\mskip1.5mu\}\;{}\<[E]%
\\
\>[3]{}\hsindent{2}{}\<[5]%
\>[5]{}\Varid{→}\;\Conid{WtSum}\;\{\mskip1.5mu \Conid{FExpr}\mskip1.5mu\}\;\{\mskip1.5mu \Conid{WtExpr}\mskip1.5mu\}\;\{\mskip1.5mu \Varid{lift⁺}\mskip1.5mu\}\;(\Varid{inj}\;\Varid{lift⁺}\;\Varid{e})\;\Varid{τ}\;{}\<[E]%
\\
\>[3]{}\hsindent{2}{}\<[5]%
\>[5]{}\Varid{→}\;\Conid{WtExpr}\;(\Varid{apply}\;\Varid{lift⁺}\;\Varid{e})\;\Varid{τ}{}\<[E]%
\\
\>[B]{}\hsindent{3}{}\<[3]%
\>[3]{}\Varid{lift-wt-array}\;\mathbin{:}\;\{\mskip1.5mu \Varid{τ}\;\mathbin{:}\;\Conid{Type}\mskip1.5mu\}\;\{\mskip1.5mu \Varid{e}\;\mathbin{:}\;[\mskip1.5mu \Conid{Array}\mskip1.5mu]\;\Conid{Expr}\mskip1.5mu\}\;{}\<[E]%
\\
\>[3]{}\hsindent{2}{}\<[5]%
\>[5]{}\Varid{→}\;\Conid{WtArray}\;\{\mskip1.5mu \Conid{FExpr}\mskip1.5mu\}\;\{\mskip1.5mu \Conid{WtExpr}\mskip1.5mu\}\;\{\mskip1.5mu \Varid{liftA}\mskip1.5mu\}\;\{\mskip1.5mu \Varid{liftℕ}\mskip1.5mu\}\;{}\<[E]%
\\
\>[5]{}\hsindent{4}{}\<[9]%
\>[9]{}(\Varid{inj}\;\Varid{liftA}\;\Varid{e})\;\Varid{τ}\;{}\<[E]%
\\
\>[3]{}\hsindent{2}{}\<[5]%
\>[5]{}\Varid{→}\;\Conid{WtExpr}\;(\Varid{apply}\;\Varid{liftA}\;\Varid{e})\;\Varid{τ}{}\<[E]%
\ColumnHook
\end{hscode}\resethooks
To define a step relation on $\mathit{Expr}$, $-\longrightarrow-$ we provide
a similar wrapping for each language component
\begin{hscode}\SaveRestoreHook
\column{B}{@{}>{\hspre}l<{\hspost}@{}}%
\column{3}{@{}>{\hspre}l<{\hspost}@{}}%
\column{5}{@{}>{\hspre}l<{\hspost}@{}}%
\column{9}{@{}>{\hspre}l<{\hspost}@{}}%
\column{E}{@{}>{\hspre}l<{\hspost}@{}}%
\>[B]{}\Keyword{data}\;\Varid{\char95 ⟶\char95 }\;\mathbin{:}\;\Conid{Expr}\;\Varid{→}\;\Conid{Expr}\;\Varid{→}\;\Conid{Set₁}\;\Keyword{where}{}\<[E]%
\\
\>[B]{}\hsindent{3}{}\<[3]%
\>[3]{}\Varid{step⁺}\;\mathbin{:}\;\{\mskip1.5mu \Varid{e}\;\mathbin{:}\;[\mskip1.5mu \Conid{Sum}\mskip1.5mu]\;\Conid{Expr}\mskip1.5mu\}\;\{\mskip1.5mu \Varid{e'}\;\mathbin{:}\;\Conid{LazyCoercion}\;\Conid{Expr}\mskip1.5mu\}\;{}\<[E]%
\\
\>[3]{}\hsindent{2}{}\<[5]%
\>[5]{}\Varid{→}\;\Varid{\char95 ⟶⁺\char95 }\;\{\mskip1.5mu \Conid{FExpr}\mskip1.5mu\}\;\{\mskip1.5mu \Varid{\char95 ⟶\char95 }\mskip1.5mu\}\;\{\mskip1.5mu \Varid{lift⁺}\mskip1.5mu\}\;\{\mskip1.5mu \Varid{liftℕ}\mskip1.5mu\}\;{}\<[E]%
\\
\>[5]{}\hsindent{4}{}\<[9]%
\>[9]{}(\Varid{inj}\;\Varid{lift⁺}\;\Varid{e})\;\Varid{e'}\;{}\<[E]%
\\
\>[3]{}\hsindent{2}{}\<[5]%
\>[5]{}\Varid{→}\;\Varid{apply}\;\Varid{lift⁺}\;\Varid{e}\;\Varid{⟶}\;\Varid{coerce}\;\Varid{e'}{}\<[E]%
\\
\>[B]{}\hsindent{3}{}\<[3]%
\>[3]{}\Varid{step}\;[\mskip1.5mu \mskip1.5mu]\;\mathbin{:}\;\{\mskip1.5mu \Varid{e}\;\mathbin{:}\;[\mskip1.5mu \Conid{Array}\mskip1.5mu]\;\Conid{Expr}\mskip1.5mu\}\;\{\mskip1.5mu \Varid{e'}\;\mathbin{:}\;\Conid{LazyCoercion}\;\Conid{Expr}\mskip1.5mu\}\;{}\<[E]%
\\
\>[3]{}\hsindent{2}{}\<[5]%
\>[5]{}\Varid{→}\;\Varid{\char95 ⟶}\;[\mskip1.5mu \mskip1.5mu]\;\anonymous \;{}\<[E]%
\\
\>[5]{}\hsindent{4}{}\<[9]%
\>[9]{}\{\mskip1.5mu \Conid{FExpr}\mskip1.5mu\}\;\{\mskip1.5mu \Varid{\char95 ⟶\char95 }\mskip1.5mu\}\;\{\mskip1.5mu \Varid{liftA}\mskip1.5mu\}\;\{\mskip1.5mu \Varid{liftℕ}\mskip1.5mu\}\;\{\mskip1.5mu \Varid{liftO}\mskip1.5mu\}\;{}\<[E]%
\\
\>[5]{}\hsindent{4}{}\<[9]%
\>[9]{}(\Varid{inj}\;\Varid{liftA}\;\Varid{e})\;\Varid{e'}\;{}\<[E]%
\\
\>[3]{}\hsindent{2}{}\<[5]%
\>[5]{}\Varid{→}\;\Varid{apply}\;\Varid{liftA}\;\Varid{e}\;\Varid{⟶}\;\Varid{coerce}\;\Varid{e'}{}\<[E]%
\ColumnHook
\end{hscode}\resethooks
The only piece remaining is to prove type preservation.  We begin in the same
way we have for each of the previous proofs using the step relation's
constructors as a guide.  The type signature should not have changed
\begin{hscode}\SaveRestoreHook
\column{B}{@{}>{\hspre}l<{\hspost}@{}}%
\column{3}{@{}>{\hspre}l<{\hspost}@{}}%
\column{E}{@{}>{\hspre}l<{\hspost}@{}}%
\>[B]{}\Varid{preservation}\;\mathbin{:}\;\{\mskip1.5mu \Varid{e}\;\Varid{e'}\;\mathbin{:}\;\Conid{Expr}\mskip1.5mu\}\;\{\mskip1.5mu \Varid{τ}\;\mathbin{:}\;\Conid{Type}\mskip1.5mu\}\;{}\<[E]%
\\
\>[B]{}\hsindent{3}{}\<[3]%
\>[3]{}\Varid{→}\;\Varid{e}\;\Varid{⟶}\;\Varid{e'}\;\Varid{→}\;\Conid{WtExpr}\;\Varid{e}\;\Varid{τ}\;\Varid{→}\;\Conid{WtExpr}\;\Varid{e'}\;\Varid{τ}{}\<[E]%
\ColumnHook
\end{hscode}\resethooks
and there are two cases $step^{+}$ and $step[]$; moreover we should expect
to merely apply preservation-* to each case, supplying the necessary lift
functions and the induction hypothesis.  This is indeed the case:
\begin{hscode}\SaveRestoreHook
\column{B}{@{}>{\hspre}l<{\hspost}@{}}%
\column{3}{@{}>{\hspre}l<{\hspost}@{}}%
\column{22}{@{}>{\hspre}l<{\hspost}@{}}%
\column{24}{@{}>{\hspre}l<{\hspost}@{}}%
\column{E}{@{}>{\hspre}l<{\hspost}@{}}%
\>[B]{}\Varid{preservation}\;(\Varid{step⁺}\;\Varid{ste})\;(\Varid{lift-wt-sum}\;\Varid{wts})\;{}\<[E]%
\\
\>[B]{}\hsindent{3}{}\<[3]%
\>[3]{}\mathrel{=}\;\Varid{preservation-Sum}\;\Varid{lift-wt-nat}\;\Varid{lift-wt-sum}\;{}\<[E]%
\\
\>[3]{}\hsindent{19}{}\<[22]%
\>[22]{}\Varid{preservation}\;\Varid{ste}\;\Varid{wts}{}\<[E]%
\\
\>[B]{}\Varid{preservation}\;(\Varid{step}\;[\mskip1.5mu \mskip1.5mu]\;\Varid{ste})\;(\Varid{lift-wt-array}\;\Varid{wta})\;{}\<[E]%
\\
\>[B]{}\hsindent{3}{}\<[3]%
\>[3]{}\mathrel{=}\;\Varid{preservation-Array}\;\Varid{lift-wt-option}\;\Varid{lift-wt-array}\;{}\<[E]%
\\
\>[3]{}\hsindent{21}{}\<[24]%
\>[24]{}\Varid{preservation}\;\Varid{ste}\;\Varid{wta}{}\<[E]%
\ColumnHook
\end{hscode}\resethooks
Having shown type preservation it is interesting to see the similarity between
how terms are shown to be welltyped and to evaluate and how the terms are
expressed in $\mu \mathit{FExpr}$.  Recall that each term in $\mathit{Expr}$ is wrapped by
a tag---given by $inj_1$ and $inj_2$---and the constructor $inn$ plays the
role of recursion.  To reiterate consider the convenience functions,
\begin{hscode}\SaveRestoreHook
\column{B}{@{}>{\hspre}l<{\hspost}@{}}%
\column{E}{@{}>{\hspre}l<{\hspost}@{}}%
\>[B]{}\Varid{nilE}\;\mathbin{:}\;\Conid{Expr}{}\<[E]%
\\
\>[B]{}\Varid{nilE}\;\mathrel{=}\;\Varid{inn}\;(\Varid{inj₂}\;(\Varid{inj₁}\;(\Varid{inj₂}\;\Varid{tt}))){}\<[E]%
\\
\>[B]{}\Varid{nat}\;\mathbin{:}\;\Conid{ℕ}\;\Varid{→}\;\Conid{Expr}{}\<[E]%
\\
\>[B]{}\Varid{nat}\;\Varid{n}\;\mathrel{=}\;\Varid{inn}\;(\Varid{inj₁}\;(\Varid{inj₁}\;(\Varid{inj₁}\;\Varid{n}))){}\<[E]%
\\
\>[B]{}\anonymous \;[\mskip1.5mu \anonymous \mskip1.5mu]\;\Varid{=\char95 }\;\mathbin{:}\;\Conid{Expr}\;\Varid{→}\;\Conid{Expr}\;\Varid{→}\;\Conid{Expr}\;\Varid{→}\;\Conid{Expr}{}\<[E]%
\\
\>[B]{}\Varid{a}\;[\mskip1.5mu \Varid{n}\mskip1.5mu]\;\mathrel{=}\;\Varid{e}\;\mathrel{=}\;\Varid{apply}\;\Varid{liftA}\;(\Varid{a}\;[\mskip1.5mu \Varid{n}\mskip1.5mu]\;\Conid{:=}\;\Varid{e}){}\<[E]%
\\
\>[B]{}\Varid{\char95 !E\char95 }\;\mathbin{:}\;\Conid{Expr}\;\Varid{→}\;\Conid{Expr}\;\Varid{→}\;\Conid{Expr}{}\<[E]%
\\
\>[B]{}\Varid{a}\;\Varid{!E}\;\Varid{n}\;\mathrel{=}\;\Varid{apply}\;\Varid{liftA}\;(\Varid{a}\;\mathbin{!}\;\Varid{n}){}\<[E]%
\\
\>[B]{}\Varid{\char95 +E\char95 }\;\mathbin{:}\;\Conid{Expr}\;\Varid{→}\;\Conid{Expr}\;\Varid{→}\;\Conid{Expr}{}\<[E]%
\\
\>[B]{}\Varid{e₁}\;\Varid{+E}\;\Varid{e₂}\;\mathrel{=}\;\Varid{apply}\;\Varid{lift⁺}\;(\Varid{e₁},\Varid{e₂}){}\<[E]%
\ColumnHook
\end{hscode}\resethooks
we may then ask: why is the term
\begin{hscode}\SaveRestoreHook
\column{B}{@{}>{\hspre}l<{\hspost}@{}}%
\column{E}{@{}>{\hspre}l<{\hspost}@{}}%
\>[B]{}\Varid{exp}\;\mathbin{:}\;\Conid{Expr}{}\<[E]%
\\
\>[B]{}\Varid{exp}\;\mathrel{=}\;(\Varid{nilE}\;[\mskip1.5mu \Varid{nat}\;\Varid{0}\mskip1.5mu]\;\mathrel{=}\;\Varid{nat}\;\Varid{1})\;\Varid{!E}\;(\Varid{nat}\;\Varid{0}\;\Varid{+E}\;\Varid{nat}\;\Varid{1}){}\<[E]%
\ColumnHook
\end{hscode}\resethooks
welltyped?  The answer given by $WtExpr$ is
\begin{hscode}\SaveRestoreHook
\column{B}{@{}>{\hspre}l<{\hspost}@{}}%
\column{3}{@{}>{\hspre}l<{\hspost}@{}}%
\column{9}{@{}>{\hspre}l<{\hspost}@{}}%
\column{17}{@{}>{\hspre}l<{\hspost}@{}}%
\column{25}{@{}>{\hspre}l<{\hspost}@{}}%
\column{30}{@{}>{\hspre}l<{\hspost}@{}}%
\column{E}{@{}>{\hspre}l<{\hspost}@{}}%
\>[B]{}\Varid{wt-exp}\;\mathbin{:}\;\Conid{WtExpr}\;\Varid{exp}\;\Conid{TOption}{}\<[E]%
\\
\>[B]{}\Varid{wt-exp}\;\mathrel{=}\;\Varid{lift-wt-array}\;(\Varid{ok-lookup}\;\Varid{wta}\;\Varid{wt+}){}\<[E]%
\\
\>[B]{}\hsindent{3}{}\<[3]%
\>[3]{}\Keyword{where}{}\<[E]%
\\
\>[3]{}\hsindent{6}{}\<[9]%
\>[9]{}\Varid{wta}\;\mathbin{:}\;\Conid{WtExpr}\;(\Varid{nilE}\;[\mskip1.5mu \Varid{nat}\;\Varid{0}\mskip1.5mu]\;\mathrel{=}\;\Varid{nat}\;\Varid{1})\;\Conid{TArray}{}\<[E]%
\\
\>[3]{}\hsindent{6}{}\<[9]%
\>[9]{}\Varid{wta}\;\mathrel{=}\;\Varid{lift-wt-array}\;{}\<[E]%
\\
\>[9]{}\hsindent{8}{}\<[17]%
\>[17]{}(\Varid{ok-ins}\;(\Varid{lift-wt-array}\;\Varid{ok-nil})\;{}\<[E]%
\\
\>[17]{}\hsindent{8}{}\<[25]%
\>[25]{}(\Varid{lift-wt-nat}\;\Varid{1})\;(\Varid{lift-wt-nat}\;\Varid{0})){}\<[E]%
\\
\>[3]{}\hsindent{6}{}\<[9]%
\>[9]{}\Varid{wt+}\;\mathbin{:}\;\Conid{WtExpr}\;(\Varid{nat}\;\Varid{0}\;\Varid{+E}\;\Varid{nat}\;\Varid{1})\;\Conid{TNat}{}\<[E]%
\\
\>[3]{}\hsindent{6}{}\<[9]%
\>[9]{}\Varid{wt+}\;\mathrel{=}\;\Varid{lift-wt-sum}\;(\Varid{ok-sum}\;{}\<[E]%
\\
\>[9]{}\hsindent{21}{}\<[30]%
\>[30]{}(\Varid{lift-wt-nat}\;\Varid{0})\;(\Varid{lift-wt-nat}\;\Varid{1})){}\<[E]%
\ColumnHook
\end{hscode}\resethooks
The $\mathit{lift{-}wt{-}*}$ functions play the same role in $\mathit{WtExpr}$ as $inn$ 
does in $\mathit{Expr}$; however rather than using the generalized approach 
of a series of disjoint sums we bundle the tag and recursion into a single 
constructor for each language component.  Evaluation displays a similar symmetry
\begin{hscode}\SaveRestoreHook
\column{B}{@{}>{\hspre}l<{\hspost}@{}}%
\column{10}{@{}>{\hspre}l<{\hspost}@{}}%
\column{E}{@{}>{\hspre}l<{\hspost}@{}}%
\>[B]{}\Varid{eval-expr}\;\mathbin{:}\;(\Varid{nilE}\;[\mskip1.5mu \Varid{nat}\;\Varid{0}\mskip1.5mu]\;\mathrel{=}\;\Varid{nat}\;\Varid{1})\;\Varid{!E}\;(\Varid{nat}\;\Varid{0}\;\Varid{+E}\;\Varid{nat}\;\Varid{1})\;{}\<[E]%
\\
\>[B]{}\hsindent{10}{}\<[10]%
\>[10]{}\Varid{⟶}\;(\Varid{nilE}\;[\mskip1.5mu \Varid{nat}\;\Varid{0}\mskip1.5mu]\;\mathrel{=}\;\Varid{nat}\;\Varid{1})\;\Varid{!E}\;\Varid{nat}\;\Varid{1}{}\<[E]%
\\
\>[B]{}\Varid{eval-expr}\;\mathrel{=}\;\Varid{step}\;[\mskip1.5mu \mskip1.5mu]\;(\Varid{stepi}\;(\Varid{step⁺}\;\Varid{stepv})){}\<[E]%
\ColumnHook
\end{hscode}\resethooks
What does the proof that $(\nilE [ \nat 0 ]= \nat 1)\,!E\,\nat 1$ is welltyped
look like?  We can compute it by invoking\begin{hscode}\SaveRestoreHook
\column{B}{@{}>{\hspre}l<{\hspost}@{}}%
\column{3}{@{}>{\hspre}l<{\hspost}@{}}%
\column{E}{@{}>{\hspre}l<{\hspost}@{}}%
\>[3]{}\Varid{preservation}\;\Varid{eval-expr}\;\Varid{wt-exp}{}\<[E]%
\ColumnHook
\end{hscode}\resethooks
which evaluates to \begin{hscode}\SaveRestoreHook
\column{B}{@{}>{\hspre}l<{\hspost}@{}}%
\column{3}{@{}>{\hspre}l<{\hspost}@{}}%
\column{4}{@{}>{\hspre}l<{\hspost}@{}}%
\column{5}{@{}>{\hspre}l<{\hspost}@{}}%
\column{E}{@{}>{\hspre}l<{\hspost}@{}}%
\>[3]{}\Varid{lift-wt-array}{}\<[E]%
\\
\>[3]{}(\Varid{ok-lookup}{}\<[E]%
\\
\>[3]{}\hsindent{1}{}\<[4]%
\>[4]{}(\Varid{lift-wt-array}{}\<[E]%
\\
\>[4]{}\hsindent{1}{}\<[5]%
\>[5]{}(\Varid{ok-ins}\;(\Varid{lift-wt-array}\;\Varid{ok-nil})\;(\Varid{lift-wt-nat}\;\Varid{1})\;(\Varid{lift-wt-nat}\;\Varid{0}))){}\<[E]%
\\
\>[3]{}\hsindent{1}{}\<[4]%
\>[4]{}(\Varid{lift-wt-nat}\;\Varid{1})){}\<[E]%
\ColumnHook
\end{hscode}\resethooks
\section{Related Work}

Independent and concurrently with our work, Delaware, et
al.~\cite{delaware} developed a solution to moduler meta-theory in
Coq.  Both their approach and ours relies on the principle of
representing data types as functors; however they have chosen to
express inductive types using Church encodings and recursive
evaluation using Mendler algebras, which requires some extra
sophistication.  Here we express types as data members of the family
of polynomial functors and apply recursive evaluation directly.  Their
approach presented is further along and has shown the important level
of robustness required by most languages while there are more
unanswered questions regarding the method presented here.

\section{Conclusion and Future Work}

We should ask if we have accomplished the goal that we set out with.
The language $\mathit{Expr}$ was given componentwise and the
boiler-plate necessary to wrap each welltyping and step relation is
minimal.  The proof of type preservation was almost immediate,
requiring only an invocation of previously defined proofs for each
component.  Moreover there is no copy and paste necessary and the
repetitive components should be automatically producable given a
sophisticated macro system where terms can be inspected by name---set
equality is non-deterministic---rather than value.

Using Agda as a proof language, although convenient, leaves the
question of consistency open.  We regard this as a minor problem and
hope that our implementation would port to Coq.  A more pertinent
problem is the definition of $\mathit{preservation}$ for
$\mathit{Expr}$---Agda is unable to prove termination and we plan to
address this soon.

The language presented is quite simple, unable to express even Euclid's
algorithm, and the method of polynomial functor's used to express $Expr$
precludes the possibility of first class function types which are critical for
functional programming.  Various solutions to this problem have been
proposed~\cite{bananas in space} and the area of recursion schemes is
rich~\cite{recursion schemes from comonads}.  A real world language calls for
much heavier sophistication, but the ideas presented here are new
and their reach is open to question and requires further exploration.


\begin{thebibliography}{1}
\bibitem{metatheory for the masses} Aydemir, et al.
  \emph{Mechanized Metatheory for the Masses: The PoplMark Challenge.} 
  In International Conference on Theorem Proving in Higher Order
  Logics (TPHOLs), August 2005.

\bibitem{malcom} Grant Malcom.
  \emph{Algebraic data types and program transformations.}
  Ph.D. thesis, Department of Computing Science, Groningen University, 1990.

\bibitem{duponcheel} Luc Duponcheel.
  \emph{Using catamorphisms, subyptes and monad transformers for writing modular
  functional interpreters.} Utrecht University, 1995.

\bibitem{dependently typed programming in agda} Ulf Norell.
  \emph{Dependently Typed Programming in Agda.}
 In Proceedings of the 4th international workshop on Types in
 language design and implementation (TLDI '09). ACM, New York, NY, USA, 1-2. 

\bibitem{bananas in space} Erik Meijer and Graham Hutton.
  \emph{Bananas in Space: Extending Fold and Unfold to Exponential Types.}
  In Proceedings of the seventh international conference on Functional
  programming languages and computer architecture (FPCA '95). ACM, New
  York, NY, USA, 324-333.

\bibitem{recursion schemes from comonads} Tarmo Uustalu, et al.
  \emph{Recursion Schemes from Comonads.}
  \emph{Nordic J. of Computing} 8, 3 (September 2001), 366-390.

\bibitem{delaware} Benjamin Delaware, Bruno C. d. S. Oliveira, Tom Schrijvers.
  \emph{Meta-Theory \`a la Carte}, unpublished manuscript, July, 2012.
\end{thebibliography}
\end{document}